\documentclass[twocolumn,amsmath,superscriptaddress,amssymb,nopacs,nofootinbib,PhysRevResearch]{revtex4-2}

\usepackage[usenames,dvips]{color}
\usepackage{graphicx}
\usepackage{dcolumn}
\usepackage{bm}
\usepackage{mathtools}
\usepackage{epstopdf}
\usepackage{esint}
\usepackage{xcolor}
\usepackage{dsfont}
\usepackage[outline]{contour}
\usepackage{nicefrac}
\usepackage{multirow}

\usepackage[colorlinks=true,citecolor=magenta,linkcolor=magenta,urlcolor=magenta]{hyperref}

\newcommand{\bea}{\begin{eqnarray}}
\newcommand{\eea}{\end{eqnarray}}
\newcommand{\bt}{\textbf}

\newcommand{\phd}{\phantom{\dag}}
\newcommand{\ph}{\phantom{.}}

\newcommand{\noi}{\noindent}
\newcommand{\no}{\nonumber}
\renewcommand{\Re}{\operatorname{Re}}
\renewcommand{\Im}{\operatorname{Im}}

\newcommand{\PK}[1]{\textcolor{black}{#1}}

\begin{document}
\def\v#1{{\bf #1}}

\title{Mechanisms for Magnetic Skyrmion Catalysis and Topological Superconductivity}

\author{Yun-Peng Huang}
\email{huangyunpeng@iphy.ac.cn}
\affiliation{Beijing National Laboratory for Condensed Matter Physics, and Institute of Physics, Chinese Academy of Sciences, Beijing 100190, China}

\author{Panagiotis Kotetes}
\email{kotetes@itp.ac.cn}
\affiliation{CAS Key Laboratory of Theoretical Physics, Institute of Theoretical Physics, Chinese Academy of Sciences, Beijing 100190, China}

\vskip 1cm

\begin{abstract}
We propose an alternative route to stabilize magnetic skyrmion textures which does not require Dzyaloshinkii-Moriya interaction, magnetic anisotropy, or an external Zeeman field. Instead, it solely relies on the emergence of flux in the system's ground state. We discuss scenarios that lead to a nonzero flux, and identify the magnetic skyrmion ground states which become accessible in its presence. Moreover, we explore the chiral superconductors obtained for the surface states of a topological crystalline insulator when two types of magnetic skyrmion crystals coexist with a pairing gap. Our work opens perspectives for engineering topological superconductivity in a minimal fashion, and promises to unearth functional topological materials and devices which may be more compatible with electrostatic control than the currently explored skyrmion-Majorana platforms.
\end{abstract}

\maketitle

\section{Introduction}\label{Sec:Intro}

Magnetic textures are advantageous ingredients for engineering Majorana quasiparticles~\cite{VolovikBook,ReadGreen,Volovik99,KitaevChain, HasanKane,QiZhang,FlensbergOppenStern,Zlotnikov,KovalevRev}, since they simultaneously break time reversal symmetry (TRS) and induce Rashba spin-orbit coupling (SOC)~\cite{MostovoyFerro,BrauneckerSOC,Choy}. One finds two distinct ways through which magnetic textures give rise to these excitations. The first relies on magnetic textures crystals (MTCs), which are magnetic textures that repeat periodically in space. The interplay of superconductivity with MTCs of the helix and skyrmion~\cite{BogdanovSkMagMet,MartinBatista,Muelhbauer,Tokura1,Tokura2,Heinze,Kawamura,LeonovMostovoy,HayamiZeroField,Wulfhekel,TripleQWiesendanger,HayamiRev,HayamiSquare} types gives rise to effective $p\,$-wave topological superconductors (TSCs) in 1D~\cite{KarstenNoSOC,Ivar,KlinovajaGraphene,NadgPerge,KotetesClassi,Nakosai,Selftuned1,Selftuned2,Selftuned3,Pientka1,Ojanen,Pientka2} and 2D~\cite{KotetesClassi,Nakosai,Mendler,WeiChen,MorrSkyrmions,MorrTripleQ,SteffensenPRR}, respectively. Recent scanning tunne\-ling microscopy (STM) measurements in helical Fe chains deposited on top of Re(0001) hint towards the presence of edge \PK{Majorana zero modes (MZMs)}~\cite{KimWiesendanger}, while the coexistence of magnetic skyrmions and superconductivity has also been demonstrated in 2D Fe/Ir magnets on Re(0001)~\cite{Kubetzka}. In the se\-cond scenario, MZMs bind to magnetic skyrmion defects~\cite{BalatskySkyrmion,Yang2016,MikeSkyrmion,Kovalev,Rex2019,Garnier2019a,Garnier2019b}, which are imposed in a chiral ferromagnet~\cite{Bogdanov,BogdanovHubert,NagaosaTokuraSkX,BogdanovPanagopoulos} coupled to a conventional superconductor. Important experimental advancements along these lines were recently made in [IrFeCoPt]/Nb hybrids~\cite{PanagopoulosSkX}. In particular, magnetic skyrmions and (anti)vortices composites were experimentally rea\-li\-zed~\cite{PanagopoulosSkyrmionVortex}, thus setting the stage for the observation of MZMs in these systems in the near future.

The above exciting experimental successes towards novel MZM platforms should be also scrutinized with respect to their degree of functionality and their potential role as the hardware for a topological quantum computer~\cite{Kitaev2003,Nayak2008}, in which, a large number of MZMs would need to be simultaneously manipulated and braided~\cite{Ivanov2001}. On one hand, Fe/Ir/Re systems typically contain a large number of domains and inhomogeneities which may render them prone to disorder, while they are not amenable to electrostatic tuning. Instead, these systems can be tailored using STM tips, whose number however is difficult to scale up. This obstacle may lead to limitations when it comes to braiding multiple MZMs. In contrast, [IrFeCoPt]/Nb hybrids appear to feature a higher degree of functionality, since it is envisaged that MZMs will be manipulated using magnetic skyrmion racetracks~\cite{Hoffman}. However, this platform presents a caveat of different nature. Such a system harbors both vortices and antivortices, which are induced by the applied magnetic field and the magnetic skyrmions, respectively. The magnetic field is required to assist the nucleation of skyrmions~\cite{BogdanovPanagopoulos}, which appear in a restricted window of the parameter space spanned by the strengths of the field, the Dzyaloshinkii-Moriya interaction (DMI)~\cite{Dzyaloshinskii,Moriya}, and the magnetic anisotropy. Therefore, here it is the me\-cha\-nism for skyrmions itself that sets constraints on the controllability and reliability of the system as a potential quantum computing platform, since the coexistence of dif\-fe\-rent types of vortices leads to complex dynamics that can hinder the implementation of braiding and thus give rise to an inherent source of noise and decoherence.

In this Manuscript, we argue that these experimental hurdles can be surpassed by con\-si\-de\-ring routes for realizing magnetic skyrmions which are free from DMI, magnetic anisotropy, and a Zeeman field. As we demonstrate, it is sufficient to violate TRS in the system by means of a nonzero ``flux'' $\vartheta$, which acts as a source for the skyrmion charge $C$. Here, the flux enters as a Lagrange multiplier which shifts the intensive free energy of the system $F$ according to $F\mapsto F-\vartheta{\cal C}$, where:
\begin{align}
{\cal C}=\frac{1}{4\pi}\int d\bm{r}\ph \bm{M}(\bm{r})\cdot\big[\partial_x\bm{M}(\bm{r})\times\partial_y\bm{M}(\bm{r})\big]\,.
\end{align}

\noi In the above, $\bm{M}(\bm{r})$ corresponds to the magnetization profile of the magnetic skyrmion crystal or defect. Since ${\cal C}$ and $C$ become nonzero simultaneously, the term $-\vartheta{\cal C}$ promotes the formation of magnetic skyrmions. For this reason, we term this mechanism as \textit{magnetic skyrmion catalysis (MSC)}. 

Notably, a number of the above preconditions are known to be fulfilled in quantum Hall systems. Indeed, magnetic skyrmions have already been theoretically predicted~\cite{Sondhi,Fertig} and experimentally observed~\cite{Barrett} in ferromagnetic quantum wells more than two decades ago. Here, however, we are not interested in situations where a strong quantizing magnetic field is present. Instead, we focus on weakly doped quasi-2D itinerant magnets with energy bands which feature a nonzero Berry curvature~\cite{Niu}. Even more, as we show here, MSC is also accessible in systems which contain band tou\-ching points and thus a singular Berry curvature. In the latter case, we find that chiral fluctuations promote magnetic skyrmions either in a spontaneous fashion, or, due to the orbital coupling to an external magnetic field. Notably, in this second scenario, MSC is mediated by a phenomenon akin to the magnetic ca\-ta\-ly\-sis known from high-energy physics~\cite{Shovkovy}, but without the formation of Landau levels. 

The magnets of interest are also assumed to feature a pairing gap, which is either inherent due to spontaneous Cooper pair formation, or, is induced by means of pro\-xi\-mi\-ty to a conventional superconductor~\cite{Sau2010Proxi,Potter2011}. In either situation, we show that the synergy of a magnetic skyrmion crystal and super\-con\-duc\-ti\-vi\-ty renders the system an effective $p+ip$ superconductor which harbors chiral Majorana edge modes~\cite{HasanKane,QiZhang}. As a consequence, such systems further set the stage for the appearance of MZMs at point-like defects, such as, in the cores of vortices introduced in the phase of the supercon\-duc\-ting~\cite{ReadGreen,Volovik99} or the magnetic texture~\cite{Steffensen2021} fields.

Our manuscript is organized as follows. In Sec.~\ref{Sec:Landau} we derive the expression of the flux $\vartheta$ which is the coefficient of the cubic term of the free energy which appears in the presence of the TRS violation. In Sec.~\ref{Sec:FluxModels} we study the behavior of $\vartheta$ for two standard models, which describe a Chern insulator and a generalized Dirac-type of model. Section~\ref{Sec:MassInduction} proposes possible routes to violate TRS and demonstrates that fluctuations promote the emergence of MSC. In Sec.~\ref{Sec:MagneticGroundStates} we identify all the possible magnetic ground states that dictate itinerant magnets with te\-tra\-go\-nal symmetry when spin-rotational symmetry is intact but TRS is broken. Afterwards, we show in Sec.~\ref{Sec:MagnetismTCI} that MSC can be realized in a Dirac model with a win\-ding of two units. Then we move on and discuss in Sec.~\ref{Sec:TSC} the implications that MSC can have on en\-gi\-nee\-ring to\-po\-lo\-gi\-cal superconductivity. This work concludes in Sec.~\ref{Sec:Conclusions} with a summary and a discussion of possible experimental platforms that appear prominent to harbor MSC. Finally, further explanations and details regarding our calculations are given in Appendices~\ref{app:AppendixA}-\ref{app:AppendixE}. 

\section{Cubic term in the Landau expansion and flux}\label{Sec:Landau}

We proceed with the exposition of the MSC. We consider that the system is free of any kind of SOC and other sources of magnetism, hence, also preserving spin rotational invariance. In addition, we assume that the magnetic instability is driven by Fermi surface ne\-sting, which in turn implies that the spin susceptibility $\chi(\bm{q})$ peaks at the star of the nesting vector $\bm{Q}$.

The band structure results from a matrix Hamiltonian $\hat{{h}}(\bm{k})\mathds{1}_{\sigma}$ where $\mathds{1}_\sigma$ is the unit matrix in spin space and $\bm{k}$ the wave vector. We develop a Landau theory for the magnetic order parameter $\bm{M}(\bm{r})$, which {\color{black}is expressed in energy units, and} results from the mean-field decoupling of a Hubbard interaction with strength $U$. The magnetic order parameter couples to the electrons through an exchange term $\propto\bm{M}(\bm{r})\cdot\bm{\sigma}$, where $\bm{\sigma}$ define the spin Pauli matrices.  

Under the above conditions, the single-particle Hamiltonian in the magnetic phase takes the following form:
\begin{align}
\hat{{h}}_{\bm{M}}(\bm{k}+\bm{q},\bm{k})=\hat{{h}}(\bm{k})\mathds{1}_{\sigma}(2\pi)^2\delta(\bm{q})+\bm{M}(\bm{q})\cdot\bm{\sigma}\,,
\end{align}

\noi where we used the Fourier transform:
\begin{align}
 \bm{M}(\bm{q})=\int d\bm{r}\ph e^{-i\bm{q}\cdot\bm{r}}\bm{M}(\bm{r})\,,
\end{align}

\noi and the Dirac delta function $\delta(\bm{q})$. The Hamiltonian is written in the formalism of second quantization as:
\begin{align}
{h}_{\bm{M}}=\int\frac{d\bm{q}}{(2\pi)^2}\int d\bm{k}\ph\bm{\psi}^{\dag}(\bm{k}+\bm{q})
\hat{{h}}_{\bm{M}}(\bm{k}+\bm{q},\bm{k})\bm{\psi}(\bm{k})\,,
\end{align}

\noi where $\bm{\psi}^\dag(\bm{k})=\big(\psi_\uparrow^\dag(\bm{k}),\,\psi_\downarrow^\dag(\bm{k})\big)$. Here, $\psi_{\uparrow,\downarrow}^\dag(\bm{k})$ creates an electron of spin projection $\uparrow,\downarrow$ and wave vector $\bm{k}$. 

To derive a Landau expansion in powers of $\bm{M}(\bm{q})$, we employ the Matsubara formalism and integrate out the electronic degrees of freedom in a standard fashion~\cite{Bruus}. The lowest term in this expansion is the quadratic one since, in contrast to the situation taking place in quantum Hall systems~\cite{Sondhi,Fertig,Barrett}, here we assume that there is no net ferromagnetic moment in the ground state when $\bm{M}(\bm{r})$ is absent. The expansion coefficient at quadratic order $\propto|\bm{M}(\bm{q})|^2$ is given by $2/U-\chi(\bm{q})$~\cite{Mendler}, and determines the leading magnetic instabilities.

TRS violation allows for a cubic term in the expansion that contains the term ${\cal L}(\bm{q},\bm{p})=\bm{M}(-\bm{q}-\bm{p})\cdot\big[\bm{M}(\bm{q})\times\bm{M}(\bm{p})\big]$, which relates to the skyrmion charge density. Similar terms have been previously discussed in a phe\-no\-me\-no\-lo\-gi\-cal fashion for chiral spin liquids~\cite{Paramekanti2017,JiangJiang,HuangSheng,Paramekanti}, and have been also shown to be inducible by a magnetic field~\cite{DiptimaSen}. However, little attention has been paid so far to the emergence of this term as a result of the topological properties of the band structure and its interplay with fluctuations. 

In the following, we provide an analytical expression for the coefficient $\vartheta$ of the respective cubic free energy term, which is written in the compact form:
\begin{align}
{F}^{(3)}=\frac{1}{4\pi}\int\frac{d\bm{q}}{(2\pi)^2}\int\frac{d\bm{p}}{(2\pi)^2}\ph\vartheta(\bm{q},\bm{p})\ph{\cal L}(\bm{q},\bm{p})\,.\no
\end{align}

\noi In the above we inserted the flux density:
\bea
\vartheta(\bm{q},\bm{p})&=&
i\int\frac{d\bm{k}}{3\pi}\ph T\sum_{{\color{black}i\omega_\nu}}\Big\{
{\rm tr}\big[\hat{{G}}({\color{black}i\omega_\nu},\bm{k})\hat{{G}}({\color{black}i\omega_\nu} ,\bm{k}-\bm{q})\no\\
&&\qquad\qquad\hat{{G}}({\color{black}i\omega_\nu},\bm{k}-\bm{q}-\bm{p})\big]-\bm{q}\leftrightarrow \bm{p}\Big\}\,,\label{eq:ThetaDensity}
\eea

\noi which is expressed in terms of the Matsubara Green function in the nonmagnetic phase:
\begin{align}
\hat{G}^{-1}({\color{black}i\omega_\nu},\bm{k})={\color{black}i\omega_\nu} -\hat{h}(\bm{k})\,.\no
\end{align}

\noi ${\color{black}\omega_\nu}$ denote fermionic Matsubara frequencies, $T$ is the temperature in energy units, and ${\rm tr}$ stands for the trace ope\-ra\-tion in the matrix space spanned by $\hat{h}(\bm{k})$.

The cubic order term ${F}^{(3)}$ takes the form $-\vartheta{\cal C}$ when the system is in the vicinity of a ferromagnetic instability. In this case, we have $\bm{q},\bm{p}\approx\bm{0}$ and thus we expand Eq.~\eqref{eq:ThetaDensity} at lowest order in $\bm{q}$ and $\bm{p}$, which leads to:
\begin{align}
\vartheta=2i\varepsilon_{znm}\int\frac{d\bm{k}}{3\pi}\ph T\sum_{{\color{black}i\omega_\nu} }
{\rm tr}\left[\hat{{G}}(k)\frac{\partial\hat{{G}}(k)}{\partial k_n}
\frac{\partial\hat{{G}}(k)}{\partial k_m}\right]
\end{align}

\noi where we set $k=({\color{black}i\omega_\nu},\bm{k})$ and introduced the totally antisymmetric Levi-Civita symbol $\varepsilon_{znm}$, with $n,m=\{x,y\}$. Repeated index summation is implied. Straightforward calculations lead to one of the key results of this work:
\begin{widetext}
\begin{align}
\vartheta=\sum_{s=0,1,2}\frac{4}{(s+1)!s!}
\sum_{\alpha}\int\frac{d\bm{k}}{\pi}\ph
i\varepsilon_{znm}\left<\partial_{k_n}\bm{u}_\alpha(\bm{k})\right|\big[\mathds{1}-\hat{{\cal P}}_{\alpha}(\bm{k})\big]
\big[\varepsilon_{\alpha}(\bm{k})-\hat{h}(\bm{k})\big]^{s-2}
\left|\partial_{k_m}\bm{u}_\alpha(\bm{k})\right>\partial_\mu^s f[\varepsilon_{\alpha}(\bm{k})-\mu],\label{eq:theta}
\end{align}
\end{widetext}

\noi where $\left|\bm{u}_{\alpha}(\bm{k})\right>$ are the eigenvectors of $\hat{h}(\bm{k})$ with corresponding energy disper\-sions $\varepsilon_\alpha(\bm{k})$ and projectors $\hat{{\cal P}}_{\alpha}(\bm{k})=\left|\bm{u}_{\alpha}(\bm{k})\right>\left<\bm{u}_{\alpha}(\bm{k})\right|$. $f(\epsilon-\mu)$ is the Fermi-Dirac distribution defined in the grand canonical ensemble and eva\-lua\-ted at an energy $\epsilon$ for a chemical potential $\mu$. 

We remark that, while in the above $\vartheta$ is given by a sum of contributions arising from each band separately, it still originates from interband-only transitions, see App.~\ref{app:AppendixA}. In fact, this is reflected in the presence of the projectors $\mathds{1}-\hat{{\cal P}}_{\alpha}(\bm{k})$. Note also that for $s=2$, the term inbetween the eigenstates can be replaced by the identity operator, and thus leads to the Berry curvature~\cite{Niu}
\begin{align}
\Omega_\alpha(\bm{k})=i\varepsilon_{znm}\big<\partial_{k_n}\bm{u}_\alpha(\bm{k})
\big|\partial_{k_m}\bm{u}_\alpha(\bm{k})\big>\,.
\end{align}

\noi of the $\alpha$-th band. The above reflects the topological origin of the flux and Eq.~\eqref{eq:theta} reveals how $\vartheta$ relates to the band structure properties.

\section{Flux in two-band models}\label{Sec:FluxModels}

We evaluate $\vartheta$ for a two band model $\hat{{h}}(\bm{k})=\bm{d}(\bm{k})\cdot\bm{\kappa}$, where $\bm{\kappa}$ denote Pauli matrices that may relate to valley or similar degrees of freedom. We use the energy dispersions $\varepsilon_\pm(\bm{k})=\pm\varepsilon(\bm{k})$ with $\varepsilon(\bm{k})=|\bm{d}(\bm{k})|$, and obtain:
\begin{align}
\vartheta=\sum_{s}^{0,1,2}\sum_\alpha^{{\color{black}\pm1}}\int\frac{d\bm{k}}{\pi}\ph\frac{(-2)^s\alpha^{1+s}\Omega(\bm{k})}{(s+1)!s![\varepsilon(\bm{k})]^{2-s}}\ph\partial_\mu^sf[-\alpha\varepsilon(\bm{k})-\mu]\label{eq:Theta2Band}
\end{align}

\noi where the quantity:
\begin{align}
\Omega(\bm{k})=\frac{1}{2}\ph\hat{\bm{d}}(\bm{k})\cdot\Big[\partial_{k_x}\hat{\bm{d}}(\bm{k})\times\partial_{k_y}\hat{\bm{d}}(\bm{k})\Big]\,
\end{align} 

\noi denotes the Berry curvature of the valence band~\cite{Niu}, with $\hat{\bm{d}}(\bm{k})=\bm{d}(\bm{k})/|\bm{d}(\bm{k})|$. In the following, we explore the behaviour of $\vartheta$ for two kinds of models. First, the case of an extended ``Dirac model'' which exhibits a single $n$-th order band touching point and, second, the Qi-Wu-Zhang model~\cite{QiWuZhang} which describes a Chern insulator or metal.

\subsection{Case of extended Dirac model}\label{sec:BCP}

We first obtain $\vartheta$ for the generalized ``Dirac model'':
\begin{align}
\bm{d}(\bm{k})=\big(\upsilon_1 k^\ell\sin(\ell\phi),m,\upsilon_2 k^\ell\cos(\ell\phi)\big),\label{eq:BCP}
\end{align}

\noi where we used polar coordinates $k_x=k\cos\phi$ and $k_y=k\sin\phi$. This model yields an $\ell$-th order band tou\-ching point for $m=0$. At zero temperature and equal ``velocities'' $\upsilon_{1,2}=\upsilon$, we find the closed form result:
\begin{align}
\vartheta_{T=0}(m)=\frac{\ell}{3}\frac{{\rm sgn}(m)}{m^2}\Big[\Theta\big(|m|-|\mu|\big)-|m|\delta\big(|m|-|\mu|\big)\Big],\label{eq:FluxZeroTemp}
\end{align}

\noi which is manifestly independent of $\upsilon$. $\Theta(\epsilon)$ denotes the Heaviside unit step function at energy $\epsilon$. We observe that the flux term is proportional to the vorticity of the band touching point. Quite remarkably, we find that given these conditions, the flux is nonzero only when the chemical potential lies within the band gap, or, \PK{when it} exactly touches the band gap edge, which yields a resonance condition arising from the delta function. 

This result appears at first sight discouraging, since the MSC seems not to be accessible for itinerant magnets in which magnetism is driven by the presence of a Fermi surface which exhibits a substantial degree of nesting. However, the observation that the flux peaks for che\-mi\-cal potential values $|\mu|$ close to $|m|$, implies that introducing a nonzero temperature can unlock MSC, since the temperature effectively broadens the energy levels. 

\begin{figure}[t!]
\centering
\includegraphics[width=0.95\columnwidth]{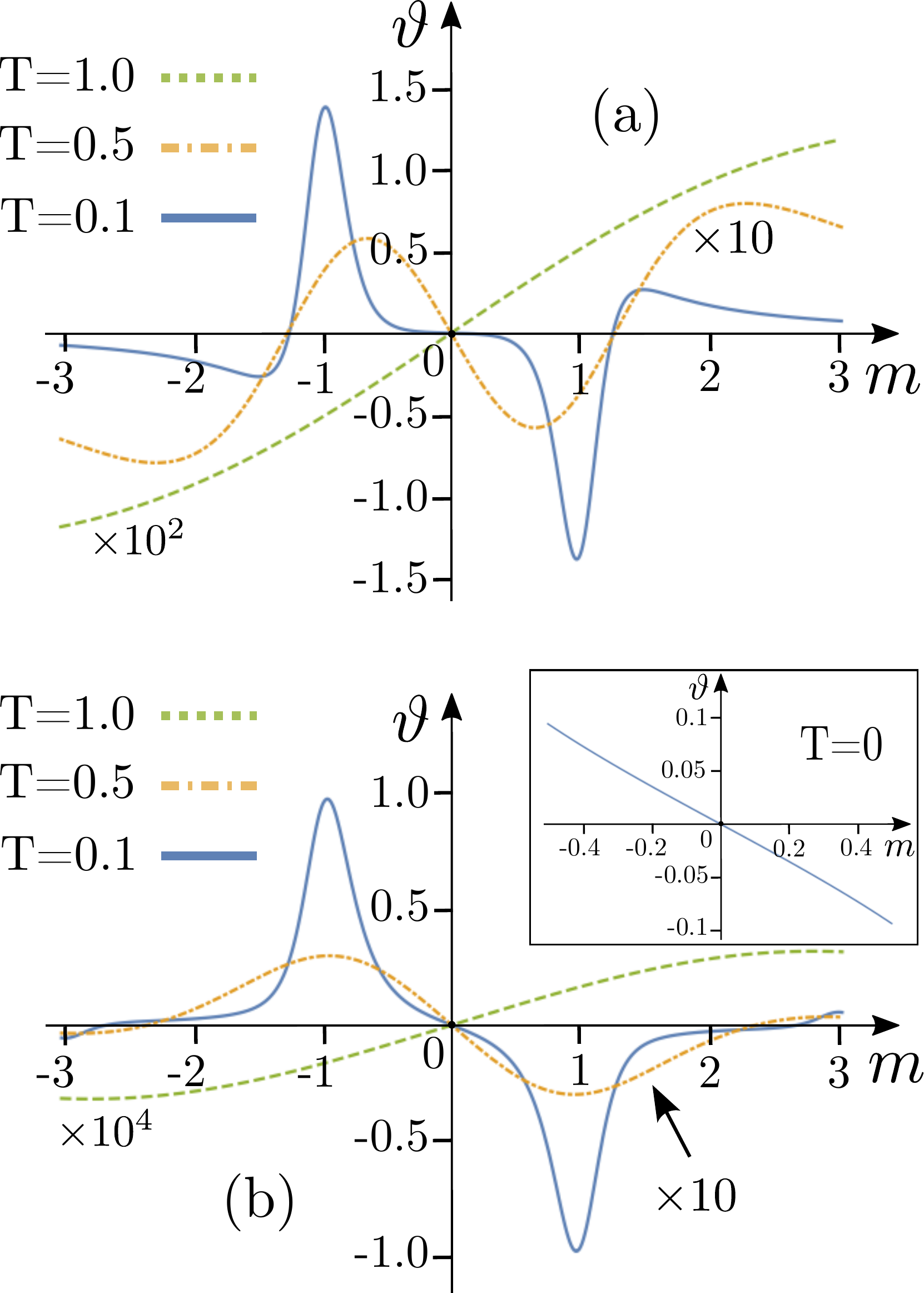}
\caption{Behavior of the coefficient $\vartheta$ as a function of tem\-pe\-ra\-tu\-re $T$ and mass/gap parameter $m$. Results for the model of the generalized model of Sec.~\ref{sec:BCP} are shown in (a), while calculations for the Chern insulator model of Sec.~\ref{sec:Chern} are shown in (b). $\vartheta$ mostly exhibits a qualitative similar beha\-vior for the two models. However, a crucial difference is that at $T=0$, $\vartheta$ is precisely zero for $|m|<|\mu|$ in the model of Eq.~\eqref{eq:BCP}, while such a restriction is not present for the Chern insulator of Eq.~\eqref{eq:ChernInsulator} as shown in the inset of (b). For the calculations we used $\mu=1$.}
\label{fig:Figure1}
\end{figure}

Indeed this is confirmed by the numerical evaluation of $\vartheta$ for nonzero temperatures shown in Fig.~\ref{fig:Figure1}(a). Therefore, when $T$ is comparable to $|m|-|\mu|$, the sy\-stem \PK{fulfills} the resonance condition and leads to nonzero flux. In the low temperature regime, the flux can be approximated by the following form:
\begin{align}
\vartheta_{\rm res}\approx-\frac{\ell}{3\pi}\frac{m}{\mu^2}\frac{T}{\big(|\mu|-|m|\big)^2+T^2}\,.\label{eq:ThetaRes} 
\end{align}

\noi The above nontrivial dependence of $\vartheta$ on the chemical potential further implies that the magnetic skyrmions contribute to the electric charge~\cite{Sondhi} since $\partial\vartheta/\partial\mu\neq0$. However, at exactly zero temperature the magnetic skyrmions do not carry electric charge since, in contrast to Ref.~\onlinecite{Sondhi}, our starting point is a paramagnetic ground state and on top of that, no SOC is present~\cite{Freimuth}. 

Besides the inclusion of a nonzero tem\-pe\-ra\-tu\-re, at this stage, it is equally important to comment on the effects of velocity anisotropy. We find that the system is practically insensitive to changing the Fermi surface properties.

\subsection{Case of Chern insulator}\label{sec:Chern}

In this pararaph we briefly discuss results for a two band Chern insulator model described by the $\bm{d}(\bm{k})$ vector:
\begin{align}
\bm{d}(\bm{k})=\big(\sin k_x,\sin k_y,\cos k_x+\cos k_y-m\big)\label{eq:ChernInsulator}
\end{align}

\noi with $\bm{k}\in[-\pi,\pi]^2$. The above model supports a topo\-lo\-gi\-cal\-ly nontrivial phases with a Chern number of a single unit $\propto{\rm sgn}(m)$ in the interval $|m|<2$.

The expression for $\vartheta$ is once again given by Eq.~\eqref{eq:Theta2Band}. Since obtaining analytical results is tedious, we restrict to a numerical investigation that reveals the salient features that cha\-rac\-te\-ri\-ze the behaviour of this system. Our calculations shown in Fig.~\ref{fig:Figure1}(b) confirm that a nonzero $\vartheta$ is obtainable even for $|\mu|>|m|$, \PK{albeit with a relatively small magnitude.} Nevertheless, the coefficient can be further boosted by considering a low nonzero temperature.

\section{Mass induction mechanisms}\label{Sec:MassInduction}

So far, the origin of the flux $\vartheta$ and the mass term $m$ in Eq.~\eqref{eq:BCP} have remained unspecified. The sole requirement to generate these, is that the energy bands see a nonzero Berry curvature. Therefore, prominent candidates appear to be narrow gap magnetic semiconductors. Preferably, these should be of a low electronic density in order to be tunable via electrostatic gating. 

MSC also arises in weakly doped semimetals which contain higher order band touching points as in Eq.~\eqref{eq:BCP} for $m=0$. However, this is possible only in the pre\-sen\-ce of fluctuations in the channel which gives rise to $m$. Indeed, the pre\-sen\-ce of such chiral fluctuations enables the system to spontaneoussly induce a nonzero $m$, since this allows for MSC to take place and, in turn, minimize the free energy by stabilizing a magnetic skyrmion crystal ground state. In the remainder of this subsection, we focus on $\ell$-th order band touching points as in Eq.~\eqref{eq:BCP}.

We make the above scenario plausible by adopting a phenomenological approach along the lines of prior works discussing fluctuations mechanisms in different contexts~\cite{KotetesPhilo,Millis,Livanas}. We focus on the mass sector of the free energy, which takes the form:
\begin{align}
F_m=\left(\frac{1}{V_m}-\chi_m\right)\frac{m^2}{2}-\vartheta(m){\cal C}-{\cal M}_z(m)B_z,\no
\end{align}

\noi where $\chi_m$ denotes the respective ``mass-mass'' susceptibility. $B_z$ denotes an out-of-plane magnetic field, which is sufficiently weak to consider the Zeeman effect fully negligible. Thus, $B_z$ couples to the electrons via the orbital effects. The latter is here introduced by means of the {\color{black}zero temperature} orbital magnetization ${\cal M}_z$~\cite{Niu,OrbitalMag}, which is a function of $m$, and is obtained from the expression:
\begin{align}
{\cal M}_z(m)=\frac{\mu}{\Phi_0}\sum_\alpha\int\frac{d\bm{k}}{2\pi}\ph \Omega_\alpha(\bm{k})f\big[\varepsilon_\alpha(\bm{k})-\mu\big]\,,\no
\end{align}

\noi where we introduced the magnetic flux quantum $\Phi_0$. Note that here we neglected the contribution of ${\cal C}$ to the orbital magnetization, since ${\cal C}$ is examined at the level of an instability analysis. For a two band model of the form Eq.~\eqref{eq:BCP}, the orbital magnetization reads:
\begin{align}
{\cal M}_z(m)=\frac{\ell m\mu}{\Phi_0}\left[\frac{\Theta\big(|m|-|\mu|\big)}{|m|}+\frac{\Theta\big(|\mu|-|m|\big)}{|\mu|}\right]\,,
\end{align}

\noi which already includes a factor of $2$ due to spin. 

In the remainder we assume low temperatures, as well as a value $|\mu|>|m|$ but for a chemical potential slightly above the conduction band bottom. These conditions allow us to approximately write ${\cal M}_z(m)\approx{\rm sgn}(\mu)\ell m/\Phi_0$. The same conditions allow us to obtain an analogous expression for $\vartheta(m)$. Specifically, we assume a thermally activated MSC and employ Eq.~\eqref{eq:ThetaRes}. After further simplifications, we consider $\vartheta\approx-\ell m /3\pi\mu^2T$.

We now move on with identifying $m$. The interaction potential $V_m$ is non-negative and thus attractive in the channel of $m$. \PK{Here}, $m$ is considered not to emerge, which is reflected in the relation $1/V_m-\chi_m>0$ assumed to hold throughout. However, \PK{the situation changes} when the se\-cond and/or the third terms are considered. The value of $m$ induced by ${\cal C}$ and/or $B_z$, that we here term $m_{\rm ind}$ is found by extremizing the free energy, i.e., $dF_m/dm=0$, which yields:
\begin{align}
m_{\rm ind}=\frac{\ell V_m}{1-V_m\chi_m}\left[\frac{{\cal C}}{3\pi\mu^2T}-{\rm sgn}(\mu)\frac{B_z}{\Phi_0}\right].\no
\end{align}

\noi We observe that even for ${\cal C}=0$, a nonzero $m$ becomes readily induced, due to the applied magnetic field. This phenomenon is a manifestation of the weak-field analog of magnetic catalysis~\cite{Shovkovy,KotetesBook}, and was previously discussed for field-induced chiral superconductors~\cite{Laughlin} and density waves~\cite{BalatskyCDDW,KotetesNernst,KotetesPhilo}. 

For the value $m_{\rm ind}$ the energy of the system is reduced. \PK{This is because a nonzero $m$ emerges in spite of the system} being detuned from \PK{an} instability in the mass channel. This can be more transparently demonstrated by introducing the field $\tilde{m}$, which is conjugate to $m$ in the statistical mechanics sense. Following Ref.~\cite{NegeleOrland}, we find:
\begin{align}
\tilde{m}=\chi_mm+\ell\left[\frac{{\cal C}}{3\pi\mu^2T}-{\rm sgn}(\mu)\frac{B_z}{\Phi_0}\right].\no
\end{align}

\noi The next step is to perform a Legendre transform which expresses the free energy in terms of $\tilde{m}$. We thus find:
\begin{align}
F_{\tilde{m}}=\frac{\big\{\tilde{m}-\ell\big[{\cal C}/3\pi\mu^2T-{\rm sgn}(\mu)B_z/\Phi_0\big]\big\}^2}{2\chi_m}-V_m\frac{\tilde{m}^2}{2}\,.\no
\end{align}

\noi By plugging in the above the induced value of $\tilde{m}$ due to ${\cal C}$ and/or $B_z$, we find that the energy is reduced by the amount: 
\begin{align}
\delta E=-\frac{\ell^2}{2}\frac{\big[{\cal C}/3\pi\mu^2T-{\rm sgn}(\mu)B_z/\Phi_0\big]^2}{1/V_m-\chi_m}\,. 
\end{align}

\noi \PK{The above} also reveals that chiral fluctuations me\-dia\-te a coupling between ${\cal C}$ and $B_z$, with the latter acting now as a source of magnetic skyrmion density.

Conclusively, we find that chiral fluctuations favor a nonzero ${\cal C}$, as they open up an additional route to reduce the free energy. Whether a nonzero ${\cal C}$ finally appears, will be decided by the remaining magnetic contributions to the free energy.

\section{Magnetic ground states in the presence of flux}\label{Sec:MagneticGroundStates}

We now proceed with discussing aspects of the magnetic instability and phase diagram for generic antiferromagnets. We assume a system with square lattice symmetry which exhibits tendency to develop magnetic ground states \PK{at the star of the ordering wave vectors} $\pm\bm{Q}_{1,2}$, with $\bm{Q}_1=Q(1,0)$ and $\bm{Q}_2=Q(0,1)$. In the presence of the nonzero flux, the cubic term also enforces ordering at wave vectors $\bm{Q}_{\pm}=\bm{Q}_1\pm\bm{Q}_2$.\footnote{Such a multi-$\bm{Q}$ magnetic ground state can be alternatively stabilized by an external field instead of flux, as it was recently experimentally observed in Ref.~\onlinecite{Takagi}.} Notably, $\bm{Q}_{\pm}$ do not belong to the star $\pm\bm{Q}_{1,2}$. Hence, if $\bm{Q}_{1,2}$ are to define the leading magnetic instability, the susceptibility at $\bm{Q}_{\pm}$ is not expected to show an equally sharp peak. Therefore, the respective magnetic orders $\bm{M}_{\bm{Q}_\pm}\equiv\bm{M}_\pm$ would not undergo an instability when considered alone\footnote{Note that in crystal lattices with trigonal and hexagonal symmetry, the cubic term involves three magnetic order parameters with all entering the magnetic skyrmion charge and determining the leading instability.}.

Let us now discuss the above rather general implementation for a square lattice in more detail. We focus on the re\-le\-vant order parameter subspace and write:
\bea
&&F_{\bm{M}_{1,2,\pm}}=F_{\bm{M}_{1,2}}+\bar{\alpha}\left(|\bm{M}_+|^2+|\bm{M}_-|^2\right)\no\\
&&-\vartheta_{1,2}\big[
\bm{M}_{+}^*\cdot\big(\bm{M}_{1}\times\bm{M}_{2}\big)-\bm{M}_{-}^*\cdot\big(\bm{M}_1\times\bm{M}_2^*\big)+{\rm c.c.}\big],\no\\
\label{eq:free_energy}
\eea

\noi where $\vartheta_{1,2}$ corresponds to $\vartheta(\bm{q},\bm{p})$ after being eva\-lua\-ted for all permutations of the momenta $\bm{q},\bm{p}=\big\{\bm{Q}_1,\bm{Q}_2,\bm{Q}_\pm\big\}$. In the above, we introduced:
\bea
F_{\bm{M}_{1,2}}&=&
\alpha\big(|\bm{M}_1|^2+|\bm{M}_2|^2\big)+\frac{\tilde{\beta}}{2}\big(|\bm{M}_1|^2+|\bm{M}_2|^2\big)^2\nonumber\\ 
&+&\frac{\beta-\tilde{\beta}}{2}\big(|\bm{M}_1^2|^2+|\bm{M}_2^2|^2\big)+(g-\tilde{\beta})|\bm{M}_1|^2|\bm{M}_2|^2\nonumber\\ 
&+&\frac{\tilde{g}}{2}\big(|\bm{M}_1\cdot\bm{M}_2|^2+|\bm{M}_1\cdot\bm{M}_2^{\ast}|^2\big)\,.\label{eq:F12}
\eea
\renewcommand{\arraystretch}{2}
\begin{table*}
\begin{centering}
\resizebox*{\textwidth}{!}{
\begin{tabular}{c|c|c}
\hline 
\hline 
Phases &Magnetic Order Parameters: $\bm{M}_1$,\, $\bm{M}_{2}$,\, $\bm{M}_{\pm}$ & Skyrmion Charge in a Magnetic Unit Cell\tabularnewline
\hline
\hline  

\multirow{1}{*}{Sk-SVC}        
& $\bm{M}_1=M\big(1,0,0\big)$,\,$\bm{M}_2=M\big(0,1,0\big)$,\,$\bm{M}_\pm=\pm\tilde{M}\big(0,0,1\big)$
& \multirow{1}{*}{$|C|=2,\qquad\tilde{M}\neq0$}\tabularnewline

\hline 
\multirow{1}{*}{Sk-MS$\|$MH}&$\bm{M}_1=M\cos\eta\big(0,0,1\big)$,\,$\bm{M}_2=M\sin\eta\big(i\sin\lambda,0,\cos\lambda\big)$,\,$\bm{M}_{\pm}=\tilde{M}\big(0,i\sin\lambda,0\big)$ 
&\multirow{1}{*}{$|C|=2,\qquad\tilde{M}\neq0 $}\tabularnewline
\hline

\multirow{2}{*}{Sk-SWC$_4$}&\multirow{1}{*}{$\bm{M}_{1}=M\big(i\cos\lambda,0,\sin\lambda\big)$, $\bm{M}_{2}=M\big(0,i\cos\lambda,\sin\lambda\big)$}
    
&\multirow{2}{*}{$
\begin{cases}
\big|C\big|=1 & |M|>|\tilde{M}\cot\lambda|\\
\big|C\big|=2 & |M|<|\tilde{M}\cot\lambda|
\end{cases}$}\tabularnewline
&\multirow{1}{*}{$\bm{M}_{\pm}=\tilde{M}\big(i\sin\lambda,\pm i\sin\lambda,\cos\lambda\big)$ }& \tabularnewline\hline 

\multirow{2}{*}{Sk-SWC$_2$}&\multirow{1}{*}{$\bm{M}_{1}=M\cos\eta\big(1,0,i\big)$, $\bm{M}_{2}=\sqrt{2}M\sin\eta\big(\cos\lambda,i\sin\lambda,0\big)$}

&\multirow{2}{*}{$
\begin{cases}
\big|C\big|=1&\big|\tilde{M}\big|<|M||\cos\eta+\sqrt{2}\sin\eta\cos\lambda|\\
\big|C\big|=2&\big|\tilde{M}\big|>|M||\cos\eta+\sqrt{2}\sin\eta\cos\lambda|
\end{cases}$}\tabularnewline
&\multirow{1}{*}{$\bm{M}_{\pm}=\tilde{M}\big(1,\pm i\cot\lambda, i\big)/2$ }& \tabularnewline\hline \hline
\end{tabular}}
\par
\end{centering}
\caption{The four magnetic leading instabilities of Eq.~\eqref{eq:free_energy} which possess a nonzero skyrmion charge in the presence of a nonzero $\gamma$. The four magnetic phases build upon the nonskyrmion analog phases which are obtained for $\gamma=0$ and are solutions of Eq.~\eqref{eq:F12}. The magnetic phases Sk-MS$||$MH and Sk-SWC$_2$ are found only as local minima (metastable) solutions.}
\label{Table:No1}
\end{table*}

All the magnetic ground states of $F_{\bm{M}_{1,2}}$, have been ana\-ly\-ti\-cal\-ly and precisely identified in Ref.~\cite{Christensen_18}. Notably, skyrmion type of crystals do not belong to the \PK{thermodynamically} stable magnetic ground states of \PK{$F_{\bm{M}_{1,2}}$}. However, among the accessible ground states of the free ener\-gy in Eq.~\eqref{eq:F12} one finds noncollinear and noncoplanar phases which can be converted into skyrmionic textures crystals. These phases correspond to the so-called spin vortex (SVC)~\cite{Lorenzana} and \PK{spin} whirl (SWC$_4$)~\cite{Christensen_18} crystal phases, and are going to  be of relevance in this work.

In contrast to Eq.~\eqref{eq:F12}, the free energy of Eq.~\eqref{eq:free_energy} in principle allows skyrmion phases by virtue of the cubic term which appears due to the TRS violation. Indeed, we verify this by obtaining an analytical solution also here, in spite of the two extra terms. In fact, an exact solution becomes possible here because the order parameters $\bm{M}_\pm$ are essentially driven by the magnetic instability at $\bm{Q}_{1,2}$, and $\bar{\alpha}$ is a non-negative number. These two conditions allow us to integrate out $\bm{M}_\pm$. The Euler-Lagrange equations of motion yield the constraints:
\begin{align}
\bm{M}_+=+\frac{\vartheta_{1,2}}{\bar{\alpha}}\bm{M}_1\times\bm{M}_2\phd\,{\rm and}\,\phd\bm{M}_-=-\frac{\vartheta_{1,2}}{\bar{\alpha}}\bm{M}_1\times\bm{M}_2^*.\label{eq:Mpm}
\end{align}

\noi Plugging the above back into Eq.~\eqref{eq:free_energy}, we find an effective free energy depending only on $\bm{M}_{1,2}$:
\begin{align}
{\cal F}_{\bm{M}_{1,2}}=F_{\bm{M}_{1,2}}-\frac{\gamma}{2}\left(|\bm{M}_1\times\bm{M}_2|^2+|\bm{M}_1\times\bm{M}_2^*|^2\right)\,,\no
\end{align}

\noi where we introduced the variable
\begin{align}
\gamma=\frac{2\vartheta_{1,2}^2}{\bar{\alpha}}\geq0\,.
\end{align}

One observes that the emergence of MSC in square lattice systems relies on a fluctuations type of mechanism, which is mediated now by the noncritical order parameters $\bm{M}_\pm$. This is a peculiarity of the given crystal symmetry group, since it does not support ordering at three or more wave vectors which add up to the null vector. In addition, it is worth noting that although ${\cal F}_{\bm{M}_{1,2}}$ alone respects TRS, magnetic skyrmion crystal ground states are now possible thanks to the additional presence of $\bm{M}_\pm$. 

To show this, we minimize ${\cal F}_{\bm{M}_{1,2}}$ with respect $\bm{M}_{1,2}$. This task is carried out in App.~\ref{app:AppendixB}. Remarkably, we find that the types of accessible magnetic ground states obtained in Ref.~\cite{Christensen_18} persist even after switching on $\gamma$. Moreover, the type of magnetic ground state which minimizes ${\cal F}_{\bm{M}_{1,2}}$, can be obtained from the magnetic phase diagram of \PK{$F_{\bm{M}_{1,2}}$}, by employing the renormalized coefficients:
\begin{align}
g\mapsto g-\gamma\quad{\rm and}\quad\tilde{g}\mapsto \tilde{g}+\gamma\,.
\end{align}

\noi A proof for the above mapping is given in App.~\ref{app:AppendixB}. With $\bm{M}_{1,2}$ at hand, we obtain $\bm{M}_\pm$ using Eq.~\eqref{eq:Mpm} and, after a Fourier transform, we determine $\bm{M}(\bm{r})$. Appendix~\ref{app:AppendixC} presents the order parameters and respective magnetization profiles of all the double-$\bm{Q}$ magnetic ground states of the free energy in Eq.~\eqref{eq:free_energy}.

Table~\ref{Table:No1} discusses the properties of the four magnetic ground states of $F_{\bm{M}_{1,2,\pm}}$, which feature a nonzero skyrmion charge for $\gamma\neq0$. These are related to the following four respective ground states of Eq.~\eqref{eq:F12}, i.e., the SVC, the SWC$_4$, as well as the MS$||$MH and the SWC$_2$. The difference between SWC$_4$ and SWC$_2$ is that the former respects and the latter violates fourfold rotational symmetry. Note that SWC$_{2,4}$ generate noncoplanar magnetic profiles, while the SVC and MS$||$MH collinear. Evenmore, we remark that only the SVC and SWC$_4$ constitute thermodynamically stable magnetic ground states of the free energy in Eq.~\eqref{eq:F12}. As a result, also Sk-MS$||$MH and Sk-SWC$_2$ have a metastable character.

We find that star\-ting from a SVC (SWC$_4$) ground state for $\gamma=0$ with $\bm{M}_\pm=\bm{0}$, can induce a Sk-SVC (Sk-SWC$_4$) phase with $|C|=2$ ($|C|=1,2$) when $\gamma$ is switched on. This is because $\bm{M}_\pm$ are now nonzero and modify $\bm{M}(\bm{r})$.

\begin{figure*}[t!]
\centering
\includegraphics[width=1\textwidth]{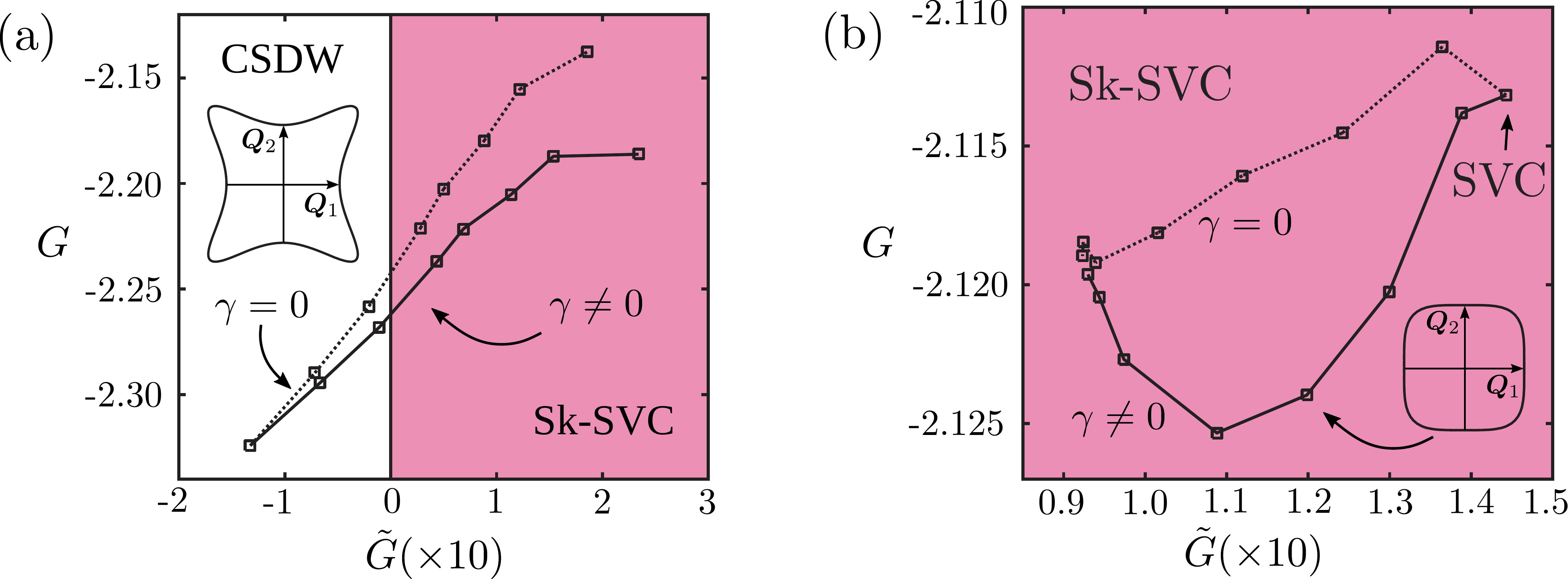}\caption{Trajectories of the leading instabilities upon varying the mass term $m$ which is added to the model of Eq.~\eqref{eq:TCIHam}. The dashed line indicates the trajectory for $\gamma=0$, i.e., without including it for the minimization of the free energy. Instead, the solid line is obtained after including the effects of a nonzero $\gamma$ which stems from a nonzero $m$. The two panels show two general Fermi surface shapes which are obtained when varying the Schr\"odinger masses $m_{1,2}$. For the Fermi surface in panel (a) [(b)], we have $m_2=0.3m_1$ [$m_2=0.7m_1$]. In all cases, the che\-mi\-cal potential is accordingly tuned so that the Fermi surface exhibits nesting with roughly the same $\bm{Q}_{1,2}$ throughout each phase diagram. When $m=0$, the CSDW (SVC) phase constitutes the leading instability in the (a) [(b)] case. Remar\-ka\-bly, for a nonzero $\gamma$ the original trajectory is pushed towards the Sk-SVC phase which now replaces the SVC phase which is obtained only for $m=0$. The axes of the phase diagram are given in terms of the variables $G=\big(g-\gamma-\tilde{\beta}\big)/|\beta-\tilde{\beta}|$ and $\tilde{G}=\big(\tilde{g}+\gamma\big)/|\beta-\tilde{\beta}|$. The values of the coefficients employed to draw the phase diagram are given in App.~\ref{app:AppendixD}.}
\label{fig:Figure2}
\end{figure*}

\section{Magnetism on the surface of a topological crystalline insulator}\label{Sec:MagnetismTCI}

In this section we demonstrate the me\-cha\-nism of MSC using as our starting point a concrete 2D model:
\begin{align}
\hat{h}_0(\bm{k})=\frac{k_xk_y}{m_1}\kappa_1+\frac{k_x^2-k_y^2}{2m_2}\kappa_3\,,\label{eq:TCIHam}
\end{align}

\noi  which leads to a single quadratic band touching point. Among other possible applications, this model also describes the protected surface states of a bulk topological crystalline insulator (TCI)~\cite{FuTCI}.

To explore the emergence of MSC for the present system, we phenomenologically introduce a mass term $m\kappa_2$ to the Hamiltonian of Eq.~\eqref{eq:TCIHam}, thus obtaining the model of Eq.~\eqref{eq:BCP} for $\ell=2$. The mass term can be induced through the mechanisms discussed in Sec.~\ref{Sec:MassInduction} and its origin is not further spe\-ci\-fied in the following. We further assume that the che\-mi\-cal potential lies ener\-ge\-ti\-cal\-ly outside the gap opened by the mass term, therefore lea\-ding to a Fermi surface. For instance, in the pa\-nels (a) and (b) of Fig.~\ref{fig:Figure2} we show two distinct type of Fermi surfaces which generally emerge for the TCI surface states. These are obtained by means of varying the che\-mi\-cal potential and/or the Schr\"odinger masses $m_{1,2}$. For suitable parameter va\-lues, the conduction band is dictated by well-nested Fermi segments. In the pre\-sen\-ce of a Hubbard interaction which is diagonal in valley $\kappa$ space, ne\-sting promotes magnetic ordering at $\bm{Q}_{1,2}$.

To identify the accessible magnetic ground states for this system, we eva\-lua\-te the Landau coefficients of the free energy in Eq.~\eqref{eq:free_energy}. We follow a standard approach that we detail in App.~\ref{app:AppendixD}. We find that a Fermi surface of the type shown in Fig.~\ref{fig:Figure2}(a), favors the stabilization of the collinear so-called charge-spin density wave (CSDW) phase~\cite{Lorenzana}. In contrast, a Fermi surface of type shown in Fig.~\ref{fig:Figure2}(b) promotes the establishment of the SVC phase. Such a tendency is in accordance with previous results for models of Fe-based systems~\cite{Christensen_Role}. There, it was shown that Fermi surfaces and interactions which are featureless in valley space, such as in Fig.~\ref{fig:Figure2}(b), promote the SVC over the CSDW phase.

These results hold both in the presence and absence of $m$. However, in the presence of the latter additional phenomena take place. When the system is originally in the SVC phase, the noncritical $\bm{M}_{\pm}$ components are also ge\-ne\-ra\-ted. Hence, a nonzero $m$ converts the SVC phase into the Sk-SVC phase with $|C|=2$. In stark contrast, the MSC does not manifest itself in the same fashion when the starting point is the CSDW phase. This is because the order parameters $\bm{M}_{1,2}$, which miminize the free energy in Eq.~\eqref{eq:F12}, are parallel in spin space. Nonetheless, for a system with CSDW instability at $m=0$, switching on $m$ can allow for a first order transition to the Sk-SVC phase with $|C|=2$.

Numerical results confirming the above are shown in Fig.~\ref{fig:Figure2}. The values for the related coefficients of the free energy in Eq.~\eqref{eq:free_energy} are given in App.~\ref{app:AppendixD}.

\section{Topological superconductivity}\label{Sec:TSC}

In this section, we continue with the investigation of the types of topological phases which become accessible in the event of coexistence of the Sk-SVC and Sk-SWC$_4$ ground states with spin-singlet $s\,$-wave super\-con\-duc\-ti\-vi\-ty. Here we have in mind situations of hybrid systems where, e.g., the surface of a bulk TCI in proximity a conventional superconductor. Nonetheless, our results can also find applicability to individual material candidates, which exhibit the microscopic coexistence of magnetism and superconductivity. Note that, while in this work we restrict to conventional pairing terms, unconventional pairing gaps also lead to topological phases in conjunction with noncollinear and noncoplanar magnetism.

\begin{figure}[t!]
\centering
\includegraphics[width=1\columnwidth]{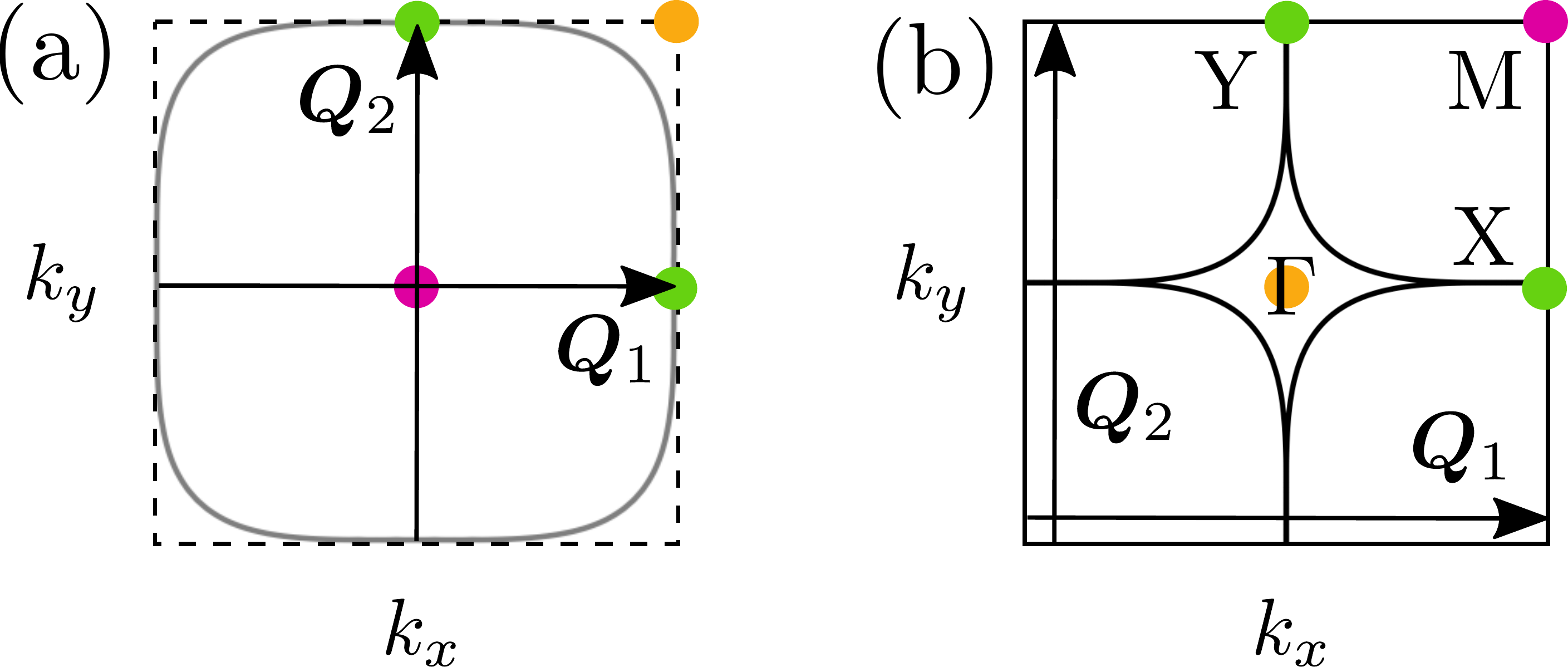}
\caption{(a) Example of a nested Fermi surface for the model of Eq.~\eqref{eq:TCIHam} with $m_1=1$, $m_2=0.7$, $m=0.038$ and $\mu=0.082$. We find that for these pa\-ra\-me\-ter values the leading magnetic instability is of the Sk-SVC type. (b) Fermi surface after downfolding to the magnetic Brillouin zone (MBZ) for the nesting vectors $\bm{Q}_{1,2}$. Note that the ${\rm \Gamma}$ point of the original Brillouin zone becomes the M point in the MBZ. On the other hand, the points X and Y of the MBZ originate from points of the Fermi surface experiencing single-$\bm{Q}$ nesting with $\bm{Q}_{1,2}$, respectively. Finally, the ${\rm \Gamma}$ point of the MBZ (orange dot) stems from a point which lies energetically away from zero, and experiences nesting with $\bm{Q}_\pm$.}
\label{fig:Figure3}
\end{figure}

A detailed exploration of the types of topological phases under various configurations of magnetic texture crystals and (un)conventional superconductivity has been presented in Ref.~\cite{SteffensenPRR}. There, it was found that the symmetries dictating the magnetic profile and the superconducting gap are pivotal for stabilizing fully-gapped and nodal TSCs, which support a variety of strong, weak and crystalline phases. In the same work, the topological properties of a SWC$_4$ phase in coexistence with conventional superconductivity were investigated. For a single band model, the symmetries dictating the SWC$_4$ magnetization profile impose that only nodal topological superconductivity is accessible in such a case. In the pre\-sen\-ce of an edge, these nodal TSCs harbor so-called bidirectional Majorana modes, which are dispersive modes that do not have a fixed sign for their group velocity. The topological mechanism that leads to these modes has a one-dimensional origin and the modes are protected by weak $\mathbb{Z}_2$ topological invariants~\cite{SteffensenPRR} which are linked to band inversions at the X and Y points of the \PK{magnetic Brillouin zone (MBZ)}.

Transitions to fully-gapped topological phases become possible in the SWC$_4$ phase only after certain types of perturbations are added~\cite{SteffensenPRR}. As we show in the following paragraphs, a nonzero $\gamma$ which stabilizes the Sk-SWC$_4$ leads to a gap in the spectrum and engineers a chiral TSC harboring a number of $|C|=1,2$ branches of chiral Majorana modes per given edge. A similar analysis for the Sk-SVC phase leads to a chiral TSC with $|C|=2$ number of chiral Majorana modes per edge. Remarkably, the number of the arising chiral edge modes is given by the respective skyrmion charge of the magnetic phase. These results qualitatively hold for any single band model that respects the square lattice symmetries and features the same type of nesting.

In the next paragraphs, we confirm the above mentioned topological phases stemming from the Sk-SVC and Sk-SWC$_4$ magnetic \PK{orders} for the model of Eq.~\eqref{eq:TCIHam}, with the remaining parameter va\-lues shown in Fig.~\ref{fig:Figure3}. Note that, while the Sk-SWC$_4$ magnetic order does not emerge as the leading magnetic instability for this model, the results of our investigation have a general character and can be applicable to systems that actually harbor this magnetic ground state.

\subsection{Bogoliubov - de Gennes formalism}

The description of topological superconductivity from magnetic texture crystals requires folding the energy dispersions of the Hamiltonian in Eq.~\eqref{eq:TCIHam} down to the  MBZ~\cite{SteffensenPRR}. For a conti\-nuum model, this process ge\-ne\-ra\-tes a band structure with an infinite number of bands. Nonetheless, only the bands crossing the Fermi level are important for inferring the topological properties of the systems under consideration. This holds as long as the magnetic and pairing gaps are sufficiently weak to allow us to restrict to only the bands contributing to the Fermi surface. In the remainder, we assume that this assumption holds. 

Within this framework, the respective many-particle Hamiltonian takes the form $H_{\rm BdG}=\frac{1}{2}\int d\bm{k}\ph\bm{{\cal X}}^\dag(\bm{k})\hat{H}_{\rm BdG}(\bm{k})\bm{{\cal X}}(\bm{k})$, where we introduced the low energy Bogoliubov - de Gennes (BdG) Hamiltonian:
\begin{widetext}
\begin{align}
\hat{H}_{\rm BdG}(\bm{k})=
\left(\begin{array}{cccc}
\hat{H}_{\rm sc}(\bm{k}-\bm{q}_+)&\bm{M}_1\cdot\bm{\sigma}&\bm{M}_2\cdot\bm{\sigma}&\bm{M}_+\cdot\bm{\sigma}\\
\bm{M}_1^*\cdot\bm{\sigma}&\hat{H}_{\rm sc}(\bm{k}+\bm{q}_-)&\bm{M}_-^*\cdot\bm{\sigma}&\bm{M}_2\cdot\bm{\sigma}\\
\bm{M}_2^*\cdot\bm{\sigma}&\bm{M}_-\cdot\bm{\sigma}&\hat{H}_{\rm sc}(\bm{k}-\bm{q}_-)&\bm{M}_1\cdot\bm{\sigma}\\
\bm{M}_+^*\cdot\bm{\sigma}&\bm{M}_2^*\cdot\bm{\sigma}&\bm{M}_1^*\cdot\bm{\sigma}&\hat{H}_{\rm sc}(\bm{k}+\bm{q}_+)
\end{array}\right)\,,\label{eq:BdG}
\end{align}

\noi which is defined in terms of the spinor:
\begin{align}
\bm{{\cal X}}^\dag(\bm{k})=\Big(\bm{\Psi}^\dag(\bm{k}-\bm{q}_+),\,\bm{\Psi}^\dag(\bm{k}+\bm{q}_-),\,\bm{\Psi}^\dag(\bm{k}-\bm{q}_-),\,\bm{\Psi}^\dag(\bm{k}+\bm{q}_+)\Big),\,
\end{align}
\end{widetext}

\noi with $\bm{q}_\pm=\bm{Q}_\pm/2$. $\bm{{\cal X}}(\bm{k})$ is in turn defined in terms of the subspinor:
\begin{align}
\bm{\Psi}^\dag(\bm{k})=\Big(\bm{\chi}_\uparrow^\dag(\bm{k}),\,\,\bm{\chi}_\downarrow^\dag(\bm{k}),\,\,\bm{\chi}_\downarrow(-\bm{k}),\,\,-\bm{\chi}_\uparrow(-\bm{k})\Big)\,,
\end{align}

\noi where $\bm{\chi}_{\uparrow,\downarrow}^\dag(\bm{k})/\bm{\chi}_{\uparrow,\downarrow}(\bm{k})$ are creation/annihilation operators of electrons with $\uparrow,\downarrow$ spin projection. Furthermore, for a fixed spin, $\bm{\chi}$ consists of two components which correspond to the valley degree of freedom $\kappa$ which leads to the valence and conduction bands of the TCI surface states of the Hamiltonian in Eq.~\eqref{eq:TCIHam}. With no loss ge\-ne\-ra\-li\-ty, the magnetic order is assumed to be diagonal in valley space.

In the general presence of a mass term $m$, the Hamiltonian $\hat{H}_{\rm sc}(\bm{k})$ describes the arising superconductor in the nonmagnetic phase, and is given as: 
\begin{align}
\hat{H}_{\rm sc}(\bm{k})=\big[\hat{h}_0(\bm{k})-\mu\big]\tau_3+m\kappa_2+\Delta\tau_1\,,\label{eq:ClassC}
\end{align}

\noi where we assumed that the surface states feel a uniform pairing gap $\Delta$ induced by means of proximity to a conventional superconductor. The Hamiltonian above is expressed in terms of the Nambu Pauli matrices $\bm{\tau}$. The latter are further supplemented by the respective identity matrix $\mathds{1}_\tau$. \PK{Note that we omit writing unit matrices and Kronecker product symbols throughout.}

The Hamiltonian $\hat{H}_{\rm sc}(\bm{k})$ belongs to symmetry class C since it is dictated by a charge conjugation symmetry $\hat{\Xi}_C^\dag\hat{H}_{\rm sc}(\bm{k})\hat{\Xi}_C=-\hat{H}_{\rm sc}(-\bm{k})$, which is effected by the ope\-ra\-tor $\hat{\Xi}_C=\tau_2\hat{{\cal K}}$. $\hat{{\cal K}}$ defines complex conjugation. Superconductors in class C exhibit to\-po\-lo\-gi\-cal\-ly nontrivial properties in 2D which are classified by a $\mathbb{Z}$ topological index~\cite{Altland1997,Schnyder2008,Kitaev2009,Ryu2010,Teo2010}. In the present situation, the emergence of topologically nontrivial behaviour is linked to the band inversion at $\bm{k}=\bm{0}$, which takes place when the criterion $|m|=\sqrt{\mu^2+\Delta^2}$ is satisfied. 

\begin{figure*}[t!]
\centering
\includegraphics[width=1\textwidth]{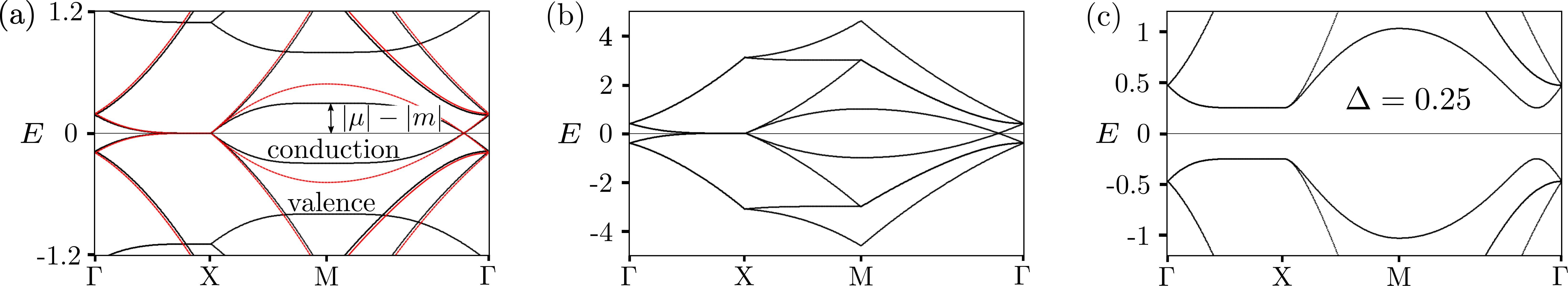}
\caption{(a) Comparison of the bare band structures ($\Delta=0$) for the full and approximate models of Eq.~\eqref{eq:ClassC} (solid black line) and~\eqref{eq:ClassCapprox} (red dashed line), after downfolding to the magnetic Brillouin zone for the nesting vectors $\bm{Q}_{1,2}$. For the model of Eq.~\eqref{eq:ClassC} we used the same parameter values employed in Fig.~\eqref{fig:Figure3}. The same values of $m_{1,2}$ were also considered for the model in Eq.~\eqref{eq:ClassCapprox}. However, in the latter case, we reduced the chemical potential by $\approx11\%$, in order to recover the same nesting properties and be in a position to compare the two band structures. Note that the full model yields twice as many bands, due to the additional presence of the valence band. As we show, the two band structures agree quite well near the Fermi level. As the energy is increased, one finds that there is a discrepancy in the vicinity of the M point, which corresponds to the $\bm{k}=\bm{0}$ point in the unfolded scheme. Therefore, as long as the pairing and magnetic gaps to be added are safely smaller than $|\mu|-|m|$, the approximate model describes quite accurately the band structure. (b) Here we depict the bare band structure obtained only from the model of Eq.~\eqref{eq:ClassCapprox} after downfolding to its respective MBZ. (c) In this panel we consider the effect due to the addition of a conventional superconducting gap on the band structure of panel (b). In all panels the energy scales are given in units of the chemical potential value $\mu=0.082$.}
\label{fig:Figure4}
\end{figure*}

\subsection{Projection onto the conduction band of the TCI surface states}

The above topological properties are however not re\-le\-vant for the cases of interest. Here, the Fermi level crosses the conduction band and the desired topological pro\-per\-ties arise from energies near the Fermi level. Hence, under the assumption that the gaps induced by the magnetic and pairing terms are much smaller than the ener\-gy difference $|\mu|-|m|$, it is eligible to project onto the conduction band of TCI surface states, and even fully discard the presence of $m$ in the resulting Hamiltonian since it leads to a minor modification of the Fermi surface shape.

The legitimacy of this approach is confirmed by comparing the resulting band structure in the MBZ of the full model and the model which only restricts to the conduction band and $m\kappa_2$ is dropped. As one observes in Fig.~\ref{fig:Figure4} for $\Delta=0$, the main difference between the full and projected model arises near the M point of the MBZ, which corresponds to the $\bm{k}=\bm{0}$ point of the unfolded $\bm{k}$ space. Nevertheless, under the weak-coupling assumption for the pairing and magnetic gaps, no band inversion takes place at M, which lies sufficiently high in energy. Therefore, using the projected model allows to qualitatively capture all the features which arise from the interplay of superconductivity and the magnetic skyrmion texture crystal near the Fermi level.

Given the above observations, in the following we approximate $\hat{H}_{\rm sc}(\bm{k})$ of Eq.~\eqref{eq:ClassC}, according to:
\begin{align}
\hat{H}_{\rm sc}(\bm{k})\approx\varepsilon(\bm{k})\tau_3+\Delta\tau_1\label{eq:ClassCapprox}
\end{align}

\noi which involves the energy dispersion:
\begin{align}
\varepsilon(\bm{k})=\sqrt{\big(k_xk_y/m_1\big)^2+\big[\big(k_x^2-k_y^2\big)/2m_2\big]^2}-\mu\,,
\end{align}

\noi where we also incorporated the chemical potential $\mu$ in it for convenience. Note, that the approximate model in Eq.~\eqref{eq:ClassCapprox} sees new nesting vectors, which have slightly modified lengths compared to the ones of the full model. Since our analysis has mainly a qualitative character, we perform our upcoming study in the MBZ defined for the nesting vectors of Eq.~\eqref{eq:ClassCapprox}. Figure~\ref{fig:Figure4}(c) depicts the low-energy band structure in the MBZ obtained from the approximate Hamiltonian in Eq.~\eqref{eq:ClassCapprox}. 

To this end, we remark that dropping $m\kappa_2$ from the Hamiltonian in order to facilitate the exploration of TSC phases stemming solely from the conduction band, does not at all imply that $m$ is fully discarded. The pre\-sence of $m$ is still captured in an indirect fashion. This is because we account for the additional emergence of the magnetic order components $\bm{M}_\pm$ which appear only when $\vartheta$ is present. We remind the reader that the latter is induced by a nonzero $m$ via the occurence of MSC.

\subsection{Compact Representation of the BdG Hamiltonian}

The Hamiltonian in Eq.~\eqref{eq:ClassCapprox}, as well as the one in Eq.~\ref{eq:BdG}, can be both written in a compact manner, which exposes their symmetries in a more transparent way. This is achieved by first introducing the Pauli matrices $\lambda_{1,2,3}$ and $\rho_{1,2,3}$ related to foldings in the $k_y$ and $k_x$ directions, respectively. These matrices correspondingly act in $\left\{\bm{k},\bm{k}+\bm{Q}_2\right\}$ and $\left\{\bm{k},\bm{k}+\bm{Q}_1\right\}$ spaces. Subsequently we consider the unitary transformation:
\begin{align}
\hat{H}_{\rm BdG}'(\bm{k})=\hat{{\cal U}}_\lambda^\dag\hat{{\cal U}}_\rho^\dag\hat{H}_{\rm BdG}(\bm{k})\hat{{\cal U}}_\rho\hat{{\cal U}}_\lambda,\label{eq:NewFrame}
\end{align}

\noi where $\hat{\cal U}_\zeta=(\zeta_3+\zeta_2)/\sqrt{2}$. Using the above, the downfolded Hamiltonian of Eq.~\eqref{eq:ClassCapprox} takes the form:
\bea
\hat{H}_{{\rm BdG;sc}}'(\bm{k})&=&\varepsilon_s(\bm{k})\tau_3+\varepsilon_x(\bm{k})\rho_2\tau_3+\varepsilon_y(\bm{k})\lambda_2\tau_3\no\\
&+&\varepsilon_{xy}(\bm{k})\lambda_2\rho_2\tau_3+\Delta\tau_1\,,
\label{eq:ClassCapproxNewFrame}
\eea

\noi where we introduced the linear combinations:
\bea
\varepsilon_s(\bm{k})&=&\frac{\varepsilon(\bm{k}-\bm{q}_+)+\varepsilon(\bm{k}+\bm{q}_-)+\varepsilon(\bm{k}-\bm{q}_-)+\varepsilon(\bm{k}+\bm{q}_+)}{4}\,\no\\
\varepsilon_x(\bm{k})&=&\frac{\varepsilon(\bm{k}-\bm{q}_+)-\varepsilon(\bm{k}+\bm{q}_-)+\varepsilon(\bm{k}-\bm{q}_-)-\varepsilon(\bm{k}+\bm{q}_+)}{4}\,\no\\
\varepsilon_y(\bm{k})&=&\frac{\varepsilon(\bm{k}-\bm{q}_+)+\varepsilon(\bm{k}+\bm{q}_-)-\varepsilon(\bm{k}-\bm{q}_-)-\varepsilon(\bm{k}+\bm{q}_+)}{4}\,\no\\
\varepsilon_{xy}(\bm{k})&=&\frac{\varepsilon(\bm{k}-\bm{q}_+)-\varepsilon(\bm{k}+\bm{q}_-)-\varepsilon(\bm{k}-\bm{q}_-)+\varepsilon(\bm{k}+\bm{q}_+)}{4}.\no
\eea

\noi The indices appearing above are chosen after atomic orbitals, in order to reflect the behavior of the above functions under the inversions $k_x\mapsto-k_x$ and/or $k_y\mapsto-k_y$.

\begin{figure*}[t!]
\centering
\includegraphics[width=\textwidth]{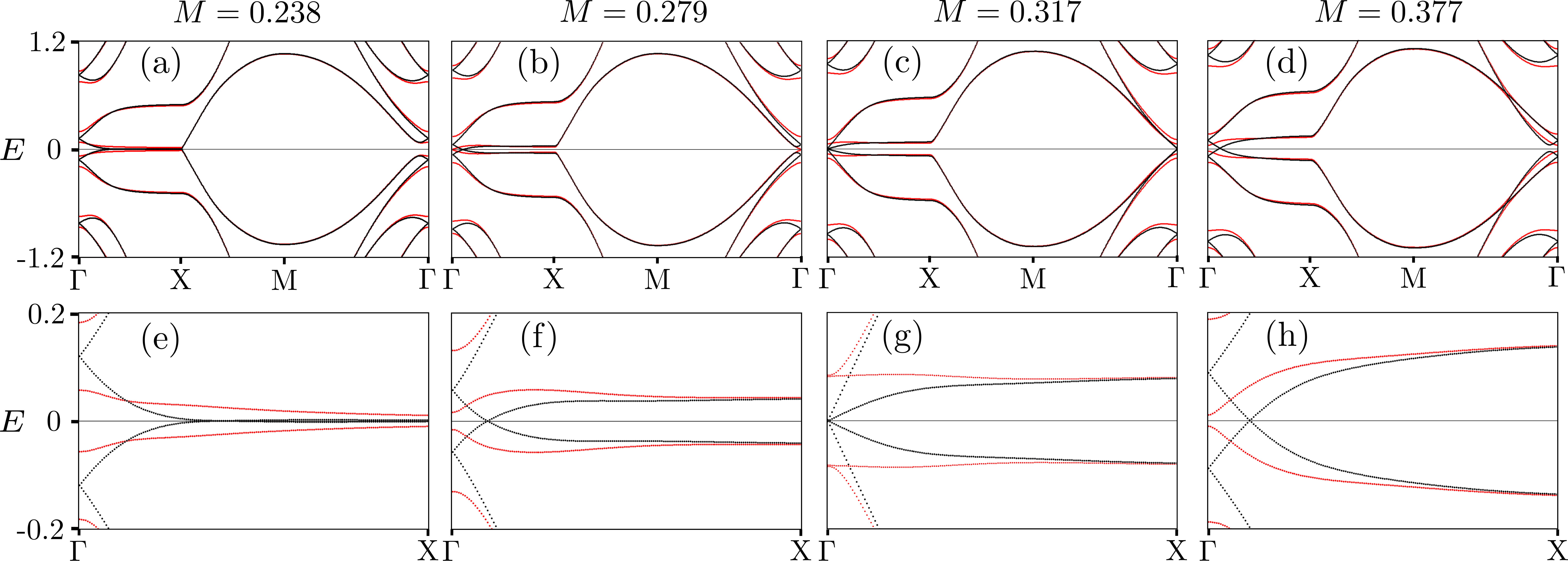}
\caption{Energy band structure for a Sk-SVC coexisting with a conventional pairing gap $\Delta=0.25$, given the bare energy dispersion shown in Fig.~\ref{fig:Figure4}(b). The latter is obtained using the parameters for $m_{1,2}$ and $\mu$ discussed in Fig.~\ref{fig:Figure3}. Red (black) color denotes dispersions in the presence (absence) of the $\bm{M}_\pm$ magnetic order parameters on top of a SVC phase. (a)-(d) and (e)-(h) show the full (top row) and a zoom-in (bottom row) of the band structure, respectively. The Sk-SVC leads to a fully-gapped spectrum with a Hamiltonian which belongs to class D$\oplus$D due to a unitary symmetry. The spectra of the two blocks are identical and allow for phases with an even number of chiral Majorana modes per edge. In all panels the energy scales are given in units of the chemical potential value $\mu=0.082$. Finally, we assumed the ratio $\tilde{M}=M/4$.}
\label{fig:Figure5}
\end{figure*}

The Hamiltonian in Eq.~\eqref{eq:ClassCapproxNewFrame} is invariant under spin rotations and commutes with $\lambda_2\rho_2$. Moreover, it is invariant under complex conjugation $\hat{{\cal K}}$ and usual time reversal operation $\hat{{\cal T}}=i\sigma_y\hat{{\cal K}}$, while it possesses a chiral symmetry effected by $\hat{\Pi}=\tau_2$ and a charge conjugation symmetry $\hat{\Xi}_D=\tau_2\sigma_y\hat{{\cal K}}$. Note that additional antiunitary symmetries can be defined by virtue of the unitary symmetry~\cite{KotetesClassi}. As we discuss in the next section, a subset of these symmetries persist inspite of the addition of magnetism.

\subsection{Topological phases: Sk-SVC ground state}

For a Sk-SVC phase, the magnetic order parameters take the form $\bm{M}_1=M(1,0,0)$, $\bm{M}_2=M(0,1,0)$, and $\bm{M}_\pm=\pm \tilde{M}(0,0,1)$. This leads to the BdG Hamiltonian in the new frame:
\begin{align}
\hat{H}_{\rm BdG}'(\bm{k})=\hat{H}_{{\rm BdG;sc}}'(\bm{k})
-M\rho_1\sigma_x-M\lambda_1\sigma_y-\tilde{M}\lambda_3\rho_3\sigma_z\,.
\label{eq:BdGSVC}
\end{align}

The above Hamiltonian preserves the earlier found antiunitary charge conjugation symmetry effected by the operator $\hat{\Xi}_D=\tau_2\sigma_y\hat{{\cal K}}$, whose presence ensures the emergence of Majorana excitations. Moreover, Eq.~\eqref{eq:BdGSVC} features a unitary symmetry generated by the operator ${\cal O}=\lambda_2\rho_2\sigma_z$. Its presence allows us to block diagonalize the Hamiltonian by means of an additional unitary transformation which renders ${\cal O}$ diagonal. We thus consider the unitary transformation effected by ${\cal S}=\big({\cal O}+\sigma_x)/\sqrt{2}$ and obtain the two-block-diagonal Hamiltonian:
\begin{align}
\hat{H}_{{\rm BdG}}''(\bm{k})=\hat{H}_{{\rm BdG;sc}}'(\bm{k})-M\rho_1\sigma_x+M\lambda_3\rho_2+\tilde{M}\lambda_1\rho_1\sigma_x\,.
\end{align}

\noi An additional unitary transformation effected by the ope\-ra\-tor ${\cal S}'=\big(\rho_3+\rho_1\sigma_x)/\sqrt{2}$ decomposes the above Hamiltonian into the two identical blocks:
\begin{align}
\hat{h}_{{\rm BdG},\sigma}(\bm{k})=\hat{h}_{{\rm BdG;sc}}'(\bm{k})-M\rho_3-M\lambda_3\rho_2+\tilde{M}\lambda_1\rho_3\,,
\label{eq:BdGSVCblock}
\end{align}

\begin{figure*}[t!]
\centering
\includegraphics[width=\textwidth]{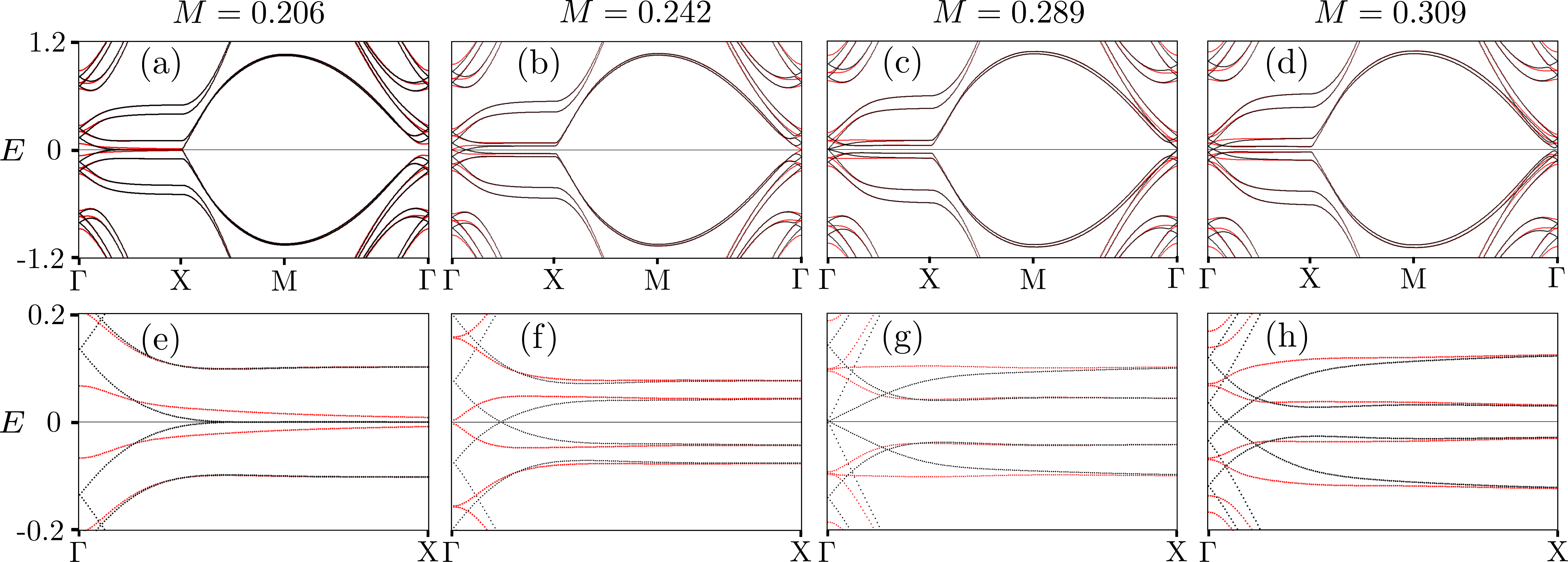}
\caption{Energy dispersions for a Sk-SWC$_4$ phase coexisting with a conventional pairing gap $\Delta=0.25$. given the bare energy dispersion shown in Fig.~\ref{fig:Figure4}(b). The latter is obtained using the parameters for $m_{1,2}$ and $\mu$ discussed in Fig.~\ref{fig:Figure3}. Red (black) color denotes dispersions in the presence (absence) of the $\bm{M}_\pm$ magnetic order parameters on top of a SVC phase. (a)-(d) and (e)-(h) show the full (top row) and a zoom-in (bottom row) of the band structure, respectively. The Sk-SWC$_4$ leads to a fully-gapped spectrum with the respective Hamiltonian belonging to class D. Note that, in contrast to the Sk-SVC case, here there is no additional twofold degeneracy. For the calculations we considered the Sk-SWC$_4$ structure shown in Table~\ref{Table:No1}, with $\lambda=\pi/4$ and $\tilde{M}=M/4$. Finally, note that in all panels the energy scales are given in units of the chemical potential value $\mu=0.082$.}
\label{fig:Figure6}
\end{figure*}

\noi which are labelled by the quantum number $\sigma$, that behaves as spin in the rotated frame. $\hat{h}_{{\rm BdG;sc}}'(\bm{k})$ is identical to $\hat{H}_{{\rm BdG;sc}}'(\bm{k})$ but restricted to a single $\sigma$ block. The appea\-ran\-ce of two identical blocks can be viewed as an emergent SO(2) spin rotational invariance in the new frame. 

In Fig.~\ref{fig:Figure5}, we present the energy bands resulting from the Hamiltonian in Eq.~\eqref{eq:BdGSVC}. We indeed confirm that the Sk-SVC phase leads to a fully-gapped twofold degenerate spectrum. The gap is opened thanks to the presence of $\bm{M}_\pm$ which are induced due the violation of TRS. 

Gap closings and reopenings are also accessible, and these happen only at the ${\rm \Gamma}$ point. The related band inversions stabilize chiral TSC phases. Indeed, we find that each one of these two blocks is dictated by a charge conjugation symmetry generated by $\hat{\Xi}=\rho_2\tau_2\hat{{\cal K}}$, which renders each Hamiltonian in class D, and allows for chiral TSCs with an integer topological invariant ${\cal N}$. Since the two blocks are identical we in fact have ${\cal N}\in2\mathbb{Z}$, and an even number of chiral Majorana modes are expected to appear per edge. 

To demonstrate the emergence of chiral TSC phases, we focus on one of the two blocks and rewrite it as:
\begin{align}
\hat{h}_{{\rm BdG},\sigma}(\bm{k})=\hat{h}_{{\rm BdG;sc}}'(\bm{k})-\sqrt{2}Me^{i\pi\lambda_3\rho_1/4}\rho_3+\tilde{M}\lambda_1\rho_3\,.\no
\end{align}

\noi As a next step, we perform a unitary transformation which transfers the phase factor $e^{i\pi\lambda_3\rho_1/4}\rho_3$ from the magnetic part of the Hamiltonian to $\hat{h}_{{\rm BdG;sc}}'(\bm{k})$. We proceed by focusing near the ${\rm \Gamma}$ point, which undergoes band inversions. We thus expand $\hat{h}_{{\rm BdG;sc}}'(\bm{k})$ up to linear order in $k_x$ and $k_y$ about $\bm{k}=\bm{0}$, and obtain the expression:
\bea
\hat{h}_{{\rm BdG},\sigma}'(\bm{k})&\approx&\sqrt{2}e^{i\pi\lambda_3\rho_1/4}\big(\upsilon_xk_x\rho_2+\upsilon_yk_y\lambda_2\big)\tau_3-\mu_{\rm \Gamma}\tau_3\no\\
&&+\Delta\tau_1-\big(\sqrt{2}M-\tilde{M}\lambda_1\big)\rho_3\,,
\label{eq:BdGSVCblockUnitary}
\eea

\noi where $\sqrt{2}\upsilon_{x,y}=\partial_{\bm{k}_{x,y}}\varepsilon_{x,y}(\bm{k})|_{\bm{k}=\bm{0}}$. $\mu_{\rm \Gamma}$ defines the effective chemical potential at ${\rm \Gamma}$. Depending on the va\-lues of $M$ and $\tilde{M}$, the band inversion at ${\rm \Gamma}$ occurs for an eigenstate with fixed projection of $\lambda_1$, i.e. $\pm1$. Therefore, it is eligible to project the above Hamiltonian onto, say, the $\lambda_1=-1$ sector, and find:
\bea
\hat{h}_{{\rm BdG},\sigma,\lambda_1=-1}'(\bm{k})&=&\big(\upsilon_xk_x\rho_2+\upsilon_yk_y\rho_1\big)\tau_3-\mu_{\rm \Gamma}\tau_3\no\\
&&+\Delta\tau_1-\big(\sqrt{2}M+\tilde{M}\big)\rho_3\,.
\label{eq:BdGSVCblockUnitaryProject}
\eea

\noi The topologically nontrivial properties of the above Hamiltonian are known from previous works~\cite{Fujimoto,SauGeneric}, and arise from phase transition mediated by the band inversion at $\bm{k}=\bm{0}$. Therefore, each class D Hamiltonian block of interest gives rise to a single branch of chiral Majorana modes at a line defect~\cite{Teo2010}.

Concluding with the analysis of this system, it is important to comment on the stability of the here-obtained chiral TSC phases with ${\cal N}=2$, since these were derived under the presence of the unitary symmetry generated by ${\cal O}$. Hence, it is particularly interesting to examine the fate of these phases in the presence of perturbations such as charge inhomogeneities and Zeeman fields, which violate ${\cal O}$. Such a study is carried out in App.~\ref{app:AppendixE} and we find that there indeed exists a certain window of stability for these phases against ${\cal O}$-symmetry-breaking perturbations.

\subsection{Topological phases: Sk-SWC$_4$ ground state}

The analysis of the Sk-SWC$_4$ can be immediately inferred from the previous investigations of topological superconductivity stemming from the SWC$_4$ phase. Spe\-ci\-fi\-cal\-ly, Ref.~\cite{SteffensenPRR} showed that only nodal TSCs become accessible, and support Majorana bidirectional modes. In fact, the nodes in the spectrum are not removable due to an emergent Kramers degeneracy arising at the ${\rm \Gamma}$ point of the band structure. As it was discussed there, lifting this degeneracy, as for instance by an out-of-plane field, permitted the appearance of chiral TSCs with a single chiral Majorana mode branch per edge.

Interestingly, the addition of the magnetization components $\bm{M}_\pm$, renders the spectrum fully gapped and achieves the removal of this Kramers degeneracy at the ${\rm \Gamma}$. Therefore, a band inversion is now possible, and allows the system to enter the topologically nontrivial phase. Given the fact that the Hamiltonian resides in class D, a single band inversion at ${\rm \Gamma}$ converts the system into a chiral TSC with ${\cal N}=1$ and a single chiral Majorana branch per line defect. 

The above statements are backed by our numerical calculations. Representative results are presented in Fig.~\ref{fig:Figure6}, where we show the pivotal character of the extra magnetic terms $\propto\bm{M}_\pm$ in splitting the degeneracy at ${\rm \Gamma}$, which appears in a system with only the SWC$_4$ magnetic phase.

\section{Conclusions and Outlook}\label{Sec:Conclusions}

We bring forward an alternative route to controllably stabilize magnetic skyrmion ground states. Our {\color{black}me\-cha\-nism that we here term as} magnetic skyrmion catalysis (MSC) is free from the requirement of Dzyaloshinkii-Moriya interaction, magnetic anisotropy and the application of an external Zeeman field. In contrast, it solely relies on time-reversal symmetry (TRS) violation, and the presence of flux $\vartheta$ in the ground state. As such, it opens perspectives for functional topological superconducting platforms that may be amenable to electrostatic or other types of control. 

{\color{black}For our exploration, we adopt a Landau-type of approach, and consider that the magnetization field $\bm{M}(\bm{r})$ is a weak perturbation to a paramagnetic ground state. Due to the violation of TRS, the Landau free energy contains a term proportional to the quantity $\int d\bm{r}\bm{M}(\bm{r})\cdot\big[\partial_x\bm{M}(\bm{r})\times\partial_y\bm{M}(\bm{r})\big]$ which is crucial for the catalysis of magnetic skyrmions. In our study the spatial varia\-tion of both the modulus $|\bm{M}(\bm{r})|$ and the orientation $\bm{M}(\bm{r})/|\bm{M}(\bm{r})|$ of the magnetization are important for the stabilization of the skyrmion magnetic ground state. This has to be contrasted with the approach adopted in a recent study~\cite{Levitov}, which considers the stabilization of smooth skyrmion textures originating from a ferromagnetic ground state. While also the authors of Ref.~\onlinecite{Levitov} independently discuss the possibility of promoting magnetic skyrmions due to Dirac points in the band structure and TRS violation, the underlying mechanism stabilizing skyrmions differs from ours.

Under the above assumptions, we provide a closed-form analytical expression for the flux $\vartheta$, which acts as a source of magnetic skyrmion charge. Using this expression}, we determine the value of this coefficient for two typical models characterized by a nonzero Berry curvature. In these cases, we study the behavior of the intrinsic flux and show that it becomes enhanced when the temperature is tuned to the energy difference between the Dirac mass $m$ and the chemical potential $\mu$. Since $m$ acts as a source of $\vartheta$ and, thus, is responsible for the MSC, we further discuss possible scenarios that allow stabilizing a nonzero $m$. An appealing possibility is to generate $m$ spontaneoussly via an attractive interaction in this channel. Nonetheless, even away from criticality, $m$ and in turn a magnetic skyrmion crystal ground state can be favored, solely by means of fluctuations. Alternatively, in the presence of fluctuations, a nonzero $m$ can be imposed by means of the orbital coupling to an applied magnetic field. {\color{black}The above results also reveal that skyrmion crystal ground states and to\-po\-lo\-gical phases generated via the MSC are compatible with the additional pre\-sen\-ce of an external orbital magnetic field and thus, in turn, with the appearance of superconducting vortices. Hence, new promising possibilities open up for pinning Majorana zero modes using skyrmion-vortex excitations, in analogy to the ones currently explored~\cite{PanagopoulosSkyrmionVortex}. Within the MSC framework, this can become possible by means of locally controlling flux in space, and generating isolated magnetic skyrmion excitations.}

In addition, we obtain the full set of possible magnetic ground states which become accessible for iti\-ne\-rant magnets with tetragonal symmetry, and preserve the full spin-rotational group, but violate TRS. This is achieved by restricting to the principal harmonics entering the magnetization profile. We find that there are two thermodynamically stable phases which acquire a nonzero skyrmion charge by means of MSC. These are the Sk-SVC and Sk-SWC$_4$ and emerge as the skyrmion variants of the spin-vortex (SVC) and spin-whirl (SWC$_4$) crystal phases which are obtained in the presence of TRS. We demonstrate the emergence of the Sk-SVC phase for a concrete extended Dirac model, which gives rise to a quadratic band crossing in the massless case, and can describe the protected surface states of a topological crystalline insulator.

Apart from the magnetic phase diagram, we also investigate the topological phases obtained for the above extended Dirac model when the two magnetic skyrmion crystal ground states mentioned above, coexist with a conventional pairing gap. This can be for instance induced to the system either by proximity to a conventional superconductor, or, spontaneously due to interactions. We find that the Sk-SVC leads to a chiral to\-po\-lo\-gi\-cal superconductor (TSC) with a topological invariant ${\cal N}=2$, which implies that chiral Majorana modes can be trapped at line defects. Instead, the Sk-SWC$_4$ allows for phases with both ${\cal N}=1,2$, depending on the details of the magnetic order parameter. Therefore, our results open alternative perspectives for engineering topological superconductivity using the mechanism of MSC, which is in principle less demanding than existing schemes for inducing magnetic skyrmion phases.

Concerning the experimental realization of MSC, we note that the SVC phase has been already found in CaKFe$_4$As$_4$~\cite{SVCexp}, while the SWC$_4$ has been theo\-re\-ti\-cal\-ly predicted~\cite{Christensen_18} for hole-doped BaFe$_2$As$_2$. Hence, Fe-based systems with coexisting magnetism and super\-con\-duc\-ti\-vi\-ty~\cite{ni08a,nandi10a,avci11,johrendt11,klauss15,wang16a} appear promising to exhibit intrinsic chiral TSCs once flux emerges. Another category of potential intrinsic TSCs is the recently discovered family of Kagome superconductors~\cite{UCDW_KVSb,muonGraf,muonKVSb,OpticalDetecCDW}. These are known to exhibit both superconductivity and a flux phase~\cite{AHE_CDW_CVSb,HiddenFluxPhaseCVSb,UCDW_RbVSb} which can lead to a nonzero Berry curvature in the energy bands~\cite{ChiralFluxKagome,Nandkishore}. Therefore, the possible \PK{additional} emergence of magnetism promises to enable the MSC and allow in turn for chiral TSC phases.

Apart from materials, further potential candidates include hybrids structures based on topological crystalline insulators~\cite{FuTCI} or graphene systems~\cite{CastroNeto,Nomura}. These are prominent for manipulating MZMs, since they can be in principle electrostatically gated. Quite remarkably, one notes that twisted bilayer graphene appears to provide all the necessary ingredients for MSC to take place, since it harbors superconductivity~\cite{CaoSC}, magnetism~\cite{CaoMag}, Chern phases~\cite{Serlin,Nuckolls}, while it has been also theoretically predicted to harbor magnetic skyrmions due to long range Coulomb interactions in its quantum Hall ferromagnetic phases~\cite{Boemerich}. 

\section*{Acknowledgements}

We wish to thank Achim Rosch for his valuable feedback on our work and for further helpful discussions. In addition, PK acknowledges funding from the National Na\-tu\-ral Science Foundation of China {\color{black}(Grant No.~12250610194).}

\newpage

\begin{appendix}

\begin{widetext}

\section{Additional details regarding the calculation of $\vartheta$}\label{app:AppendixA}

\noi The expansion of Eq.~\eqref{eq:ThetaDensity} for $\bm{q},\bm{p}\approx\bm{0}$, yields:
\bea
\vartheta(\bm{q},\bm{p})=i\int\frac{d\bm{k}}{3\pi}\ph T\sum_{{\color{black}i\omega_\nu}}
\bigg\{{\rm tr}\left[\hat{G}({\color{black}i\omega_\nu},\bm{k})\frac{\partial\hat{G}({\color{black}i\omega_\nu},\bm{k})}{\partial k_x}
\frac{\partial\hat{G}({\color{black}i\omega_\nu},\bm{k})}{\partial k_y}\right]-\frac{\partial}{\partial k_x}\leftrightarrow\frac{\partial}{\partial k_y}\bigg\}\big(q_xp_y-q_yp_x\big)\,.\no
\eea

\noi Since the integrand of $F^{(3)}$ is antisymmetric in $\bm{p}\leftrightarrow\bm{q}$, the factor $q_xp_y-q_yp_x$ can be replaced by $2q_xp_y$. Therefore, near a ferromagnetic instability the cubic term reads $F^{(3)}=-{\vartheta}{\cal C}$ where:
\bea
\vartheta=2i\int\frac{d\bm{k}}{3\pi}\ph T\sum_{{\color{black}i\omega_\nu}}\left\{{\rm tr}\left[\hat{G}({\color{black}i\omega_\nu},\bm{k})\frac{\partial\hat{G}({\color{black}i\omega_\nu},\bm{k})}{\partial k_x}
\frac{\partial\hat{G}({\color{black}i\omega_\nu},\bm{k})}{\partial k_y}\right]-\frac{\partial}{\partial k_x}\leftrightarrow\frac{\partial}{\partial k_y}\right\}\,.\no
\eea

\noi Using the relation for a general Green function $\hat{G}\hat{G}^{-1}=1\Rightarrow\partial\hat{G}=-\hat{G}(\partial\hat{G}^{-1})\hat{G}=\hat{G}(\partial\hat{{\cal H}})\hat{G}$ we obtain:
\begin{align}
{\rm tr}
\left[\hat{G}({\color{black}i\omega_\nu},\bm{k})
\frac{\partial\hat{G}({\color{black}i\omega_\nu},\bm{k})}{\partial k_x}\frac{\partial\hat{G}({\color{black}i\omega_\nu},\bm{k})}{\partial k_y}\right]={\rm tr}
\left[\hat{G}({\color{black}i\omega_\nu},\bm{k})\hat{G}({\color{black}i\omega_\nu},\bm{k})\frac{\partial\hat{h}(\bm{k})}{\partial k_x}\hat{G}({\color{black}i\omega_\nu},\bm{k})
\hat{G}({\color{black}i\omega_\nu},\bm{k})\frac{\partial\hat{h}(\bm{k})}{\partial k_y}\hat{G}({\color{black}i\omega_\nu},\bm{k})\right].\no
\end{align}

\noi We now proceed by expressing the Green functions with the help of projectors. At this point we consider that the system is described by a band structure with energy dispersions $\varepsilon_\alpha(\bm{k})$. For calculating ${\vartheta}$ we will additionaly introduce a chemical potential $\mu$ and work in the grand-canonical ensemble. Under these conditions we obtain:
\begin{align}
{\rm tr}\left[
\hat{G}({\color{black}i\omega_\nu},\bm{k})\frac{\partial\hat{G}({\color{black}i\omega_\nu},\bm{k})}{\partial k_x}\frac{\partial\hat{G}({\color{black}i\omega_\nu},\bm{k})}{\partial k_y}\right]=\sum_{\alpha,\beta}
{\rm tr}
\left[\frac{\hat{{\cal P}}_{\alpha}(\bm{k})}{[{\color{black}i\omega_\nu}+\mu-\varepsilon_\alpha(\bm{k})]^3}\frac{\partial\hat{h}(\bm{k})}{\partial k_x}\frac{\hat{{\cal P}}_{\beta}(\bm{k})}{[{\color{black}i\omega_\nu}+\mu-\varepsilon_\beta(\bm{k})]^2}\frac{\partial\hat{h}(\bm{k})}{\partial k_y}\right]\,.\no
\end{align}

\noi Notably, in the above sum only the $\alpha\neq\beta$ contributions will survive due to the antisymmetry under the exchange of the derivatives. Therefore we have:
\begin{align}
{\vartheta}=2i\int\frac{d\bm{k}}{3\pi}\sum_{\alpha\neq\beta}T\sum_{{\color{black}i\omega_\nu}}\frac{1}{[{\color{black}i\omega_\nu}+\mu-\varepsilon_\alpha(\bm{k})]^3}\frac{1}{[{\color{black}i\omega_\nu}+\mu-\varepsilon_\beta(\bm{k})]^2}\left\{
{\rm tr}\left[
\hat{{\cal P}}_{\alpha}(\bm{k})\frac{\partial\hat{h}(\bm{k})}{\partial k_x}
\hat{{\cal P}}_{\beta}(\bm{k})\frac{\partial\hat{h}(\bm{k})}{\partial k_y}\right]
-\frac{\partial}{\partial k_x}\leftrightarrow\frac{\partial}{\partial k_y}\right\}\no
\end{align}

\noi and after symmetrizing, we find:
\begin{align}
{\vartheta}=i\int\frac{d\bm{k}}{3\pi}\sum_{\alpha\neq\beta}T\sum_{{\color{black}i\omega_\nu}}\frac{\varepsilon_\alpha(\bm{k})-\varepsilon_\beta(\bm{k})}{\big\{[{\color{black}i\omega_\nu}+\mu-\varepsilon_\alpha(\bm{k})][{\color{black}i\omega_\nu}+\mu-\varepsilon_\beta(\bm{k})]\big\}^3}\left\{{\rm tr}\left[\hat{{\cal P}}_{\alpha}(\bm{k})\frac{\partial\hat{h}(\bm{k})}{\partial k_x}
\hat{{\cal P}}_{\beta}(\bm{k})\frac{\partial\hat{h}(\bm{k})}{\partial k_y}\right]
-\frac{\partial}{\partial k_x}\leftrightarrow\frac{\partial}{\partial k_y}\right\}\,.\no
\end{align}

\noi We can further simplicify the above expression using relations satisfied by the eigenstates of the system $\left|\bm{u}_\alpha(\bm{k})\right>$. We specifically obtain:
\begin{align}
{\vartheta}=i\int\frac{d\bm{k}}{3\pi}\sum_{\alpha\neq\beta}T\sum_{{\color{black}i\omega_\nu}}\frac{[\varepsilon_\alpha(\bm{k})-\varepsilon_\beta(\bm{k})]^3}
{\big\{[{\color{black}i\omega_\nu}+\mu-\varepsilon_\alpha(\bm{k})][{\color{black}i\omega_\nu}+\mu-\varepsilon_\beta(\bm{k})]\big\}^3}\Big[
\big<\partial_{k_x}\bm{u}_\alpha(\bm{k})\big|\bm{u}_\beta(\bm{k})\big>\big<\bm{u}_\beta(\bm{k})\big|\partial_{k_y}\bm{u}_\alpha(\bm{k})\big>-\partial_{k_x}\leftrightarrow\partial_{k_y}\Big].\no
\end{align}

\noi After the Matsubara summation we obtain
\bea
{\vartheta}
&=&i\int\frac{d\bm{k}}{6\pi}\sum_{\alpha\neq\beta}
\big<\partial_{k_x}\bm{u}_\alpha(\bm{k})\big|\bm{u}_\beta(\bm{k})\big>
\Bigg\{12\frac{f[\varepsilon_\alpha(\bm{k})-\mu]-f[\varepsilon_\beta(\bm{k})-\mu]}
{\big[\varepsilon_\alpha(\bm{k})-\varepsilon_\beta(\bm{k})\big]^2}-6\frac{f'[\varepsilon_\alpha(\bm{k})-\mu]+f'[\varepsilon_\beta(\bm{k})-\mu]}{\varepsilon_\alpha(\bm{k})-\varepsilon_\beta(\bm{k})}\no\\
&&\qquad\qquad+f''[\varepsilon_\alpha(\bm{k})-\mu]-f''[\varepsilon_\beta(\bm{k})-\mu]\Bigg\}\big<\bm{u}_\beta(\bm{k})\big|\partial_{k_y}\bm{u}_\alpha(\bm{k})\big>-\partial_{k_x}\leftrightarrow\partial_{k_y}\,.\no
\eea

\noi To obtain the result of Eq.~\eqref{eq:theta}, we first exchange the band indices $\alpha\leftrightarrow\beta$ in the terms containing the Fermi-Dirac distributions of the $\beta$ bands. This gives rise to an overall factor of $2$ to the terms containing the Fermi-Dirac distribution of the $\alpha$ bands. Afterwards, we eliminate the dependence on the $\beta$ bands by expressing $\varepsilon_\beta(\bm{k})$ in the operator form $\hat{h}(\bm{k})$. One finally arrives at Eq.~\eqref{eq:theta} is finally by expressing the sum containing the various orders of derivatives of the Fermi-Dirac distribution in a compact fashion. 

\section{Minimization of the magnetic free energy for a nonzero flux}\label{app:AppendixB}

In this appendix we minimize the free energy in Eq.~\eqref{eq:free_energy} with respect $\bm{M}_{1,2}$ and identify also the constrained $\bm{M}_\pm$. Our analysis follows the spirit of Ref.~\cite{Christensen_18}. At quadratic order of the expansion of the free energy for $\bm{M}_{1,2}$, a degeneracy and enhanced symmetry emerges. Hence, this allows us to adopt the parametrization $\bm{M}_{1}=M\cos\eta\ph\hat{\bm{n}}_1$ and $\bm{M}_{2}=M\sin\eta\ph\hat{\bm{n}}_2$, with $|\hat{\bm{n}}_{1,2}|^2=1$ and $\eta\in[0,\pi/2]$. Thus the Landau functional reads:
\bea
&&{\cal F}_{\bm{M}_{1,2}}=\alpha M^2+\frac{\tilde{\beta}}{2}M^4
+\frac{\beta-\tilde{\beta}}{2}\big(\cos^4\eta|\hat{\bm{n}}_1^2|^2+\sin^4\eta|\hat{\bm{n}}_2^2|^2\big)M^4\no\\
&&\phd\qquad+\left[g-\tilde{\beta}+\tilde{g}\ph\frac{|\hat{\bm{n}}_1\cdot\hat{\bm{n}}_2|^2+|\hat{\bm{n}}_1\cdot\hat{\bm{n}}_2^*|^2}{2}
-\gamma\ph\frac{|\hat{\bm{n}}_1\times\hat{\bm{n}}_2|^2+|\hat{\bm{n}}_1\times\hat{\bm{n}}_2^*|^2}{2}\right]\sin^2(2\eta)\frac{M^4}{4}\,.\qquad
\eea

The presence of a nonzero $\gamma$ solely affects the double-$\bm{Q}$ phases. There we focus on these. In order to extremize the free energy functional in the case of double-$\bm{Q}$ phases ($\eta\neq0,\pi/2$) it is convenient first to write the explicit form of the two complex spin-vectors as $\hat{\bm{n}}_s=(|a_s|e^{i\zeta_s},|b_s|e^{i\xi_s},|c_s|e^{i\omega_s})$ with $s=1,2$. By virtue of translational invariance we can arbitrarily choose the overall phase factor of each complex vector. This allows us to set $\omega_s=0$ ($s=1,2$). We can now proceed by further decomposing the complex vectors $\hat{\bm{n}}_{s}$ ($s=1,2$) into real and imaginary parts, as in the previous paragraph. After taking into account translational invariance and by also exploiting the SO(3) spin-rotational invariance allowing us to fix the spin-orientation of one of the four real vectors ($\hat{\bm{n}}_{s,\Re}\,,\hat{\bm{n}}_{s,\Im}$) we can write: 
\bea
\hat{\bm{n}}_{1,\Re}&=&\cos\lambda_1\left(0\,,0\,,1\right)\,,\qquad\no\\
\hat{\bm{n}}_{1,\Im}&=&\sin\lambda_1\left(\cos\chi\,,\sin\chi\,,0\right)\,,\no\qquad\\
\hat{\bm{n}}_{2,\Re}&=&\cos\lambda_2\left(\sin\rho\cos\omega\,,\sin\rho\sin\omega\,,\cos\rho\right)\,,\qquad\no\\
\hat{\bm{n}}_{2,\Im}&=&\sin\lambda_2\left(\cos\phi\,,\sin\phi\,,0\right)\,.\no
\eea

\noi For compactness, we now define the coefficients:  
\bea
P=\frac{|\hat{\bm{n}}_1\cdot\hat{\bm{n}}_2|^2+|\hat{\bm{n}}_1\cdot\hat{\bm{n}}_2^*|^2}{2}&=&
(\hat{\bm{n}}_{1,\Re}\cdot\hat{\bm{n}}_{2,\Re})^2+(\hat{\bm{n}}_{1,\Im}\cdot\hat{\bm{n}}_{2,\Im})^2+(\hat{\bm{n}}_{1,\Re}\cdot\hat{\bm{n}}_{2,\Im})^2+
(\hat{\bm{n}}_{1,\Im}\cdot\hat{\bm{n}}_{2,\Re})^2\,,\no\\
K=\frac{|\hat{\bm{n}}_1\times\hat{\bm{n}}_2|^2+|\hat{\bm{n}}_1\times\hat{\bm{n}}_2^*|^2}{2}&=&
(\hat{\bm{n}}_{1,\Re}\times\hat{\bm{n}}_{2,\Re})^2+(\hat{\bm{n}}_{1,\Im}\times\hat{\bm{n}}_{2,\Im})^2+(\hat{\bm{n}}_{1,\Re}\times\hat{\bm{n}}_{2,\Im})^2+
(\hat{\bm{n}}_{1,\Im}\times\hat{\bm{n}}_{2,\Re})^2\,.\no
\eea

\noi Let us now calculate a number of terms that appear in the free energy
\bea
|\hat{\bm{n}}_1^2|^2&=&\cos^2(2\lambda_1)\,,\no\\
|\hat{\bm{n}}_2^2|^2&=&\cos^2(2\lambda_2)+\sin^2(2\lambda_2)\sin^2\rho\cos^2\delta_2\,,\no\\
(\hat{\bm{n}}_{1,\Re}\cdot\hat{\bm{n}}_{2,\Re})^2&=&\cos^2\lambda_1\cos^2\lambda_2\cos^2\rho\,,\no\\
(\hat{\bm{n}}_{1,\Im}\cdot\hat{\bm{n}}_{2,\Im})^2&=&\sin^2\lambda_1\sin^2\lambda_2\cos^2\delta_1\,,\no\\
(\hat{\bm{n}}_{1,\Re}\cdot\hat{\bm{n}}_{2,\Im})^2&=&0\,,\no\\
(\hat{\bm{n}}_{1,\Im}\cdot\hat{\bm{n}}_{2,\Re})^2&=&\sin^2\lambda_1\cos^2\lambda_2\sin^2\rho\cos^2\delta_3\,,\no\\
(\hat{\bm{n}}_{1,\Re}\times\hat{\bm{n}}_{2,\Re})^2&=&\cos^2\lambda_1\cos^2\lambda_2-(\hat{\bm{n}}_{1,\Re}\cdot\hat{\bm{n}}_{2,\Re})^2\,,\no\\
(\hat{\bm{n}}_{1,\Im}\times\hat{\bm{n}}_{2,\Im})^2&=&\sin^2\lambda_1\sin^2\lambda_2-(\hat{\bm{n}}_{1,\Im}\cdot\hat{\bm{n}}_{2,\Im})^2\,,\no\\
(\hat{\bm{n}}_{1,\Re}\times\hat{\bm{n}}_{2,\Im})^2&=&\cos^2\lambda_1\sin^2\lambda_2-(\hat{\bm{n}}_{1,\Re}\cdot\hat{\bm{n}}_{2,\Im})^2\,,\no\\
(\hat{\bm{n}}_{1,\Im}\times\hat{\bm{n}}_{2,\Re})^2&=&\sin^2\lambda_1\cos^2\lambda_2-(\hat{\bm{n}}_{1,\Im}\cdot\hat{\bm{n}}_{2,\Re})^2\,,\no
\eea

\noi where we set $\delta_1=\phi-\chi$, $\delta_2=\omega-\phi$ and $\delta_3=\chi-\omega$. From the above we find the relation $K=1-P$. This implies that one can go back to the start and rewrite the Landau functional as:
\bea
{\cal F}_{\bm{M}_{1,2}}&=&\alpha M^2+\frac{\tilde{\beta}}{2}M^4+\frac{\beta-\tilde{\beta}}{2}\big(\cos^4\eta|\hat{\bm{n}}_1^2|^2+\sin^4\eta|\hat{\bm{n}}_2^2|^2\big)M^4\no\\
&&+\left[g-\gamma-\tilde{\beta}+\left(\tilde{g}+\gamma\right)\frac{|\hat{\bm{n}}_1\cdot\hat{\bm{n}}_2|^2+|\hat{\bm{n}}_1\cdot\hat{\bm{n}}_2^*|^2}{2}
\right]\sin^2(2\eta)\frac{M^4}{4}\,.
\eea

\noi Thus, we find that the presence of a nonzero $\chi$, renormalizes the following parameters at $\chi=0$ according to the replacement $g\mapsto g-\gamma$ and $\tilde{g}\mapsto\tilde{g}+\gamma$. 

\section{Magnetic ground states in the presence of nonzero flux}\label{app:AppendixC}

Based on the above results, we find that all the magnetic ground states obtained in the presence of a nonzero flux, can be obtained using the ground states for zero flux discussed in Ref.~\cite{Christensen_18}. First of all, we note that the single-$\bm{Q}$ phase, i.e., the magnetic stripe and helix do not induce nonzero $\bm{M}_\pm$. Therefore, their expression remain as previous. From the double-$\bm{Q}$ phases, we also find that neither the CSDW induces the two additional magnetization components, since it is a collinear phase. The expression for the remaining six magnetic ground states are presented in Table~\ref{Table:No2}.

\begin{table*}[h!]
\begin{centering}
\resizebox*{\textwidth}{!}{
\begin{tabular}{c|c}
\hline
Phases 
& $\bm{M}_{1}\big(\bm{r}\big)$, $\bm{M}_{2}\big(\bm{r}\big)$, $\bm{M}_{\pm}\big(\bm{r}\big)$\tabularnewline
\hline 
\hline 
\multirow{1}{*}{Sk-SVC}

&\multirow{1}{*}{$\bm{M}_1=M\big(\cos\big(\bm{Q}_1\cdot\bm{r}\big),0,0\big)$, $\bm{M}_2=M\big(0,\cos\big(\bm{Q}_2\cdot\bm{r}\big),0\big)$, $\bm{M}_\pm=\pm\tilde{M}\cos\big(\bm{Q}_\pm\cdot\bm{r}\big)\big(0,0,1\big)$}\tabularnewline
\hline 

\multirow{2}{*}{Sk-MS$\|$MH} 
& $\bm{M}_{1}=\left(0,0,M\cos\eta\cos\left(\bm{Q}_{1}\cdot\bm{r}\right)\right)$, $\bm{M}_{2}=M\sin\eta\big(\sin\lambda\sin\big(\bm{Q}_2\cdot\bm{r}\big),0,\cos\lambda\cos\big(\bm{Q}_2\cdot\bm{r}\big)\big)$ \tabularnewline
    
& $\bm{M}_{\pm}=\big(0,\tilde{M}\sin\lambda\sin\eta\cos\eta\sin\big(\bm{Q}_\pm\cdot\bm{r}\big),0\big)$\tabularnewline

\hline 

\multirow{2}{*}{$\mathrm{Sk-SWC_{4}}$}
& $\bm{M}_{1}=M\left(\cos\lambda\sin\left(\bm{Q}_{1}\cdot\bm{r}\right),0,\sin\lambda\cos\left(\bm{Q}_{1}\cdot\bm{r}\right)\right)$ 
, $\bm{M}_{2}=M\left(0, \cos\lambda\sin\left(\bm{Q}_{2}\cdot\bm{r}\right),\sin\lambda\cos\left(\bm{Q}_{2}\cdot\bm{r}\right)\right)$ \tabularnewline
    
& $\bm{M}_{\pm}=\tilde{M}\left(\sin\lambda\sin\left(\bm{Q}_\pm\cdot\bm{r}\right),\pm\sin\lambda\sin\left(\bm{Q}_\pm\cdot\bm{r}\right),\cos\lambda\cos\left(\bm{Q}_\pm\cdot\bm{r}\right)\right)$\tabularnewline

\hline 

\multirow{2}{*}{$\mathrm{Sk-SWC_{2}}$ }
& $\bm{M}_{1}=M\cos\eta\left(\cos\left(\bm{Q}_{1}\cdot\bm{r}\right),0,\sin\left(\bm{Q}_{1}\cdot\bm{r}\right)\right)$, $\bm{M}_{2}=\sqrt{2}M\sin\eta\left(\cos\lambda\cos\left(\bm{Q}_{2}\cdot\bm{r}\right),\sin\lambda\sin\left(\bm{Q}_{2}\cdot\bm{r}\right),0\right)$ \tabularnewline
    
& $\bm{M}_{\pm}=\tilde{M}\left(\cos\left(\bm{Q}_\pm\cdot\bm{r}\right),\pm\cot\lambda\sin\left(\bm{Q}_\pm\cdot\bm{r}\right),\sin\left(\bm{Q}_\pm\cdot\bm{r}\right)\right)/2$\tabularnewline

\hline 
\multirow{2}{*}{$\widetilde{\mathrm{MS\perp MH}}$}
& $\bm{M}_{1}=\left(0,0,M\cos\eta\cos\left(\bm{Q}_{1}\cdot\bm{r}\right)\right)$, $\bm{M}_{2}=M\sin\eta\left(\sin\left(\bm{Q}_{2}\cdot\bm{r}\right),\cos\left(\bm{Q}_{2}\cdot\bm{r}\right),0\right)/\sqrt{2}$ \tabularnewline

& $\bm{M}_{\pm}=\tilde{M}\sin\eta\cos\eta\left(\mp\cos\left(\bm{Q}_\pm\cdot\bm{r}\right),\sin\left(\bm{Q}_\pm\cdot\bm{r}\right),0\right)$\tabularnewline
\hline 

\multirow{2}{*}{$\widetilde{\mathrm{DPMH}}$}
& $\bm{M}_{1}=M\left(\sin\left(\bm{Q}_{1}\cdot\bm{r}\right),0,\cos\left(\bm{Q}_{1}\cdot\bm{r}\right)\right)/\sqrt{2}$ 
, $\bm{M}_{2}=M\left(\sin\left(\bm{Q}_{2}\cdot\bm{r}\right),0,\cos\left(\bm{Q}_{2}\cdot\bm{r}\right)\right)/\sqrt{2}$ \tabularnewline
    
& $\bm{M}_{\pm}=\tilde{M}\left(0,\left(i\mp i\right)\sin\left(\bm{Q}_\pm\cdot\bm{r}\right),0\right)$\tabularnewline
\hline 

\hline
\end{tabular}
}
\par
\end{centering}
\caption{The coordinate-space magnetization profile $\bm{M}(\bm{r})$, for the six double-$\bm{Q}$ magnetic phases of Ref.~\cite{Christensen_18}, which become modified in the presence of a nonzero flux in the free energy of Eq.~\eqref{eq:free_energy}.}\label{Table:No2}
\end{table*}

\section{Landau free energy coefficients and phase diagram details}\label{app:AppendixD}

We can explicitly calculate the coefficients for the Landau free energy Eq.~\eqref{eq:free_energy} given the concrete model we consider in Eq.~\eqref{eq:TCIHam} after phenomenologically adding a mass term $m$. The Hamiltonian of the system can be written as $\hat{h}(\bm{k})=\bm{d}(\bm{k})\cdot\bm{\kappa}$ with the help of the Pauli matrix vector $\bm{\kappa}$ and $\bm{d}(\bm{k})=\left(k_xk_y/m_1,m,\big(k_x^2-k_y^2\big)/2m_2\right)$. We introduce the bare Matsubara Green function (for convenience we employ a slighly modified notation below):
\begin{equation}
\hat{G}\left({\color{black}i\omega_\nu},\bm{k}\right)=\frac{1}{{\color{black}i\omega_\nu}-\hat{h}(\bm{k})}
=\frac{1}{2}\underset{\alpha=\pm}{\sum}\frac{\mathds{1}+\alpha\hat{\bm{d}}(\bm{k})\cdot\bm{\kappa}}{{\color{black}i\omega_\nu}-\varepsilon_{\alpha}(\bm{k})}
=\underset{\alpha=\pm}{\sum}\frac{\hat{\hat{\mathcal{P}}}_{\alpha}(\bm{k})}{{\color{black}i\omega_\nu}-\varepsilon_{\alpha}(\bm{k})}\\
=G_{{\color{black}i\omega_\nu},\bm{k}}^{\alpha}\hat{\mathcal{P}}_{\alpha}(\bm{k}).
\end{equation}

\noi Here, $\hat{\mathcal{P}}_{\alpha}(\bm{k})=\big[\mathds{1}+\alpha\hat{\bm{d}}(\bm{k})\cdot\bm{\kappa}\big)]/2$
is the projector representing the valley degree of freedom, $\hat{\bm{d}}(\bm{k})\equiv\bm{d}(\bm{k})/\left|\bm{d}(\bm{k})\right|$, and $G_{{\color{black}i\omega_\nu},\bm{k}}^{\alpha}=\left({\color{black}i\omega_\nu}-\varepsilon_{\alpha}(\bm{k})\right)^{-1}$. It is helpful to define the following quantities:
\begin{equation}
A_{\bm{k}_{1},\alpha_{1};\bm{k}_{2},\alpha_{2};\cdots\cdots;\bm{k}_{n} \alpha_{n}}^{\left(n\right)}=\mathrm{tr}\big\{ \hat{\mathcal{P}}_{\alpha_{1}}\left(\bm{k}_{1}\right)\hat{\mathcal{P}}_{\alpha_{2}}\left(\bm{k}_{2}\right)\cdots\cdots\hat{\mathcal{P}}_{\alpha_{n}}\left(\bm{k}_{n}\right)\big\}.
\end{equation}

The coefficients of the Landau free energy are given by taking the derivatives of it with respect to the order parameters $\bm{M}_{\bm{q}}$. The free energy functional is expressed in terms of the perturbation in operator form $\hat{V}=\bm{M}_{\bm{q}}\cdot\bm{\sigma}$:
\begin{equation}
F=F_0-T \,\mathrm{Tr}\ln\left(1-\hat{G}\hat{V}\right)={F}_0+T \,\sum_{n=1}^\infty\frac{1}{n}\mathrm{Tr}\big(\hat{G}\hat{V}\big)^n\equiv F_0+\sum_{n=1}^{\infty}F^{(n)}\,.
\end{equation}

\subsection*{The Second Order Terms}

The second order contribution to the free energy $F^{\left(2\right)}$ is written as:
\begin{equation}
{F}^{\left(2\right)}=\alpha\left(\left|\bm{M}_1\right|^{2}+\left|\bm{M}_2\right|^{2} \right)+\bar{\alpha}\left(|\bm{M}_+|^2+|\bm{M}_-|^2\right)\,.
\end{equation}

\noi The coefficients $\alpha$ and $\bar{\alpha}$ contain the spin susceptibility, which is given by the expression: 
\begin{align}
\chi(\bm{q})=-\frac{2T}{N}\sum_{{\color{black}i\omega_\nu},\bm{k}}\sum_{\alpha,\beta=\pm}\,A_{\bm{k},\alpha;\bm{k}+\bm{q},\beta}^{\left(2\right)}G_{{\color{black}i\omega_\nu},\bm{k}}^{\alpha}G_{{\color{black}i\omega_\nu},\bm{k}+\bm{q}}^{\beta}
\end{align}

\noi with the projector $A_{\bm{k},\alpha;\bm{k}',\beta}^{(2)}=\big(1+\alpha\beta\hat{\bm{d}}_{\bm{k}}\cdot\hat{\bm{d}}_{\bm{k}'}\big)/2$. Considering a Hubbard interaction with strength $U$, yields the coefficients $\alpha=2/U-\chi(\bm{Q}_{1,2}),$ and $\bar{\alpha}=2/U-\chi(\bm{Q}_{\pm})$. Specifically, the interaction strength $U$ is set to meet the Stoner criterion $2=U\chi(\bm{Q}_{1,2})$, and thus we have: $\bar{\alpha}=\chi(\bm{Q}_{1,2})-\chi(\bm{Q}_{\pm})$.

\subsection*{The Third Order Terms}

The third-order free energy is written as:
\begin{align}
F^{(3)}=&\frac{T}{3 N}\sum_{\bm{k}}\mathrm{Tr}\left\{\hat{G}\left({\color{black}i\omega_\nu},\bm{k}\right)\bm{M}_{\bm{k}-\bm{k}'}\cdot\bm{\sigma}\hat{G}\left({\color{black}i\omega_\nu},\bm{k}'\right)\bm{M}_{\bm{k}'-\bm{k}''}\cdot\bm{\sigma}\hat{G}\left({\color{black}i\omega_\nu},\bm{k}''\right)\bm{M}_{\bm{k}''-\bm{k}}\cdot\bm{\sigma}\right\} \nonumber\\
=&-\vartheta_{1,2}\left[\bm{M}^{*}_{+}\cdot\left(\bm{M}_1\times\bm{M}_2\right)-\bm{M}^{*}_{-}\cdot\left(\bm{M}_1\times\bm{M}^*_2\right)+\mathrm{c.c}\right].
\end{align}

\noi We can start with calculating the projector:
\begin{align}
A_{\bm{k},s;\bm{k}',s';\bm{k}'',s''}^{\left(3\right)}= & \mathrm{Tr}\left\{ \hat{\mathcal{P}}_{s}(\bm{k})\hat{\mathcal{P}}_{s'}\left(\bm{k}'\right)\hat{\mathcal{P}}_{s''}\left(\bm{k}''\right)\right\} \nonumber\\
= &\frac{1}{4}\left(1+ss'\hat{\bm{d}}_{\bm{k}}\cdot\hat{\bm{d}}_{\bm{k}'}+ss''\hat{\bm{d}}_{\bm{k}}\cdot\hat{\bm{d}}_{\bm{k}''}+s's''\hat{\bm{d}}_{\bm{k}'}\cdot\hat{\bm{d}}_{\bm{k}''}+i\epsilon_{ij\ell}ss's''\hat{\bm{d}}_{\bm{k}}^{i}\hat{\bm{d}}_{\bm{k}'}^{j}\hat{\bm{d}}_{\bm{k}''}^{\ell}\right).
\end{align}

For term of $\bm{M}_{+}^{*}\cdot\left(\bm{M}_{1}\times\bm{M}_{2}\right)$,
the expression of the coefficient is generated by considering the functional derivatives:
\bea
\vartheta_{1,2}&=&-\frac{2iT}{N}\sum_{{\color{black}i\omega_\nu},\bm{k}}\underset{s,s',s''=\pm}{\sum}\Bigg(A_{\bm{k},s;\bm{k}-\bm{Q}_{2},s';\bm{k}+\bm{Q}_{1},s''}^{\left(3\right)}G_{{\color{black}i\omega_\nu},\bm{k}}^{s}G_{{\color{black}i\omega_\nu},\bm{k}-\bm{Q}_{2}}^{s'}G_{{\color{black}i\omega_\nu},\bm{k}+\bm{Q}_{1}}^{s''}\no\\&&\qquad\qquad\qquad\qquad\phd-A_{\bm{k},s;\bm{k}-\bm{Q}_{1},s';\bm{k}+\bm{Q}_{2},s''}^{\left(3\right)}G_{{\color{black}i\omega_\nu},\bm{k}}^{s}G_{{\color{black}i\omega_\nu},\bm{k}-\bm{Q}_{1}}^{s'}G_{{\color{black}i\omega_\nu},\bm{k}+\bm{Q}_{2}}^{s''}\Bigg)
\eea

\noi where we note that: 
\begin{align}
A_{\bm{k},s;\bm{k}',s';\bm{k}'',s''}^{\left(3\right)}= & A_{\bm{k}',s';\bm{k}'',s'';\bm{k},s;}^{\left(3\right)}=A_{\bm{k}'',s'';\bm{k},s;\bm{k}',s'}^{\left(3\right)}.
\end{align}

\noi If TRS holds for the system, the projector $A^{\left(3\right)}$ is real and the coeffcient $\vartheta_{1,2}$ is zero. When the TRS of the system breaks, the triple product term of $\hat{\bm{d}}_{\bm{k}}$ in the $A_{\bm{k},s;\bm{k}',s^{\prime};\bm{k}'',s^{\prime\prime}}^{\left(3\right)}$ becomes nonzero and leads to an imaginary factor. Notice that here the imaginary parts of $A_{\bm{k},s;\bm{k}-\bm{Q}_{2},s^{\prime};\bm{k}+\bm{Q}_{1},s^{\prime\prime}}^{\left(3\right)}$ and $A_{\bm{k},s;\bm{k}-\bm{Q}_{1},s^{\prime};\bm{k}+\bm{Q}_{2},s^{\prime\prime}}^{\left(3\right)}$ have the opposite sign when integrated, so the expression of the coefficients can be simplified as:
\begin{equation}
\vartheta_{1,2}=\frac{2T}{N}\sum_{{\color{black}i\omega_\nu},\bm{k}}\underset{s,s',s''=\pm}{\sum} 2 \Im\left(A_{\bm{k},s;\bm{k}-\bm{Q}_{2},s';\bm{k}+\bm{Q}_{1},s''}^{\left(3\right)}G_{{\color{black}i\omega_\nu},\bm{k}}^{s}G_{{\color{black}i\omega_\nu},\bm{k}-\bm{Q}_{2}}^{s'}G_{{\color{black}i\omega_\nu},\bm{k}+\bm{Q}_{1}}^{s''}\right).
\end{equation}

\subsection*{The Fourth Order Terms}

\noi Similarly, we can write down the fourth order free energy of the system:
\bea
F^{\left(4\right)}&=&\frac{T}{4 N}\sum_{{\color{black}i\omega_\nu},\bm{k}}\mathrm{Tr}\Big\{\hat{G}({\color{black}i\omega_\nu},\bm{k})\bm{M}_{\bm{k}-\bm{k}'}\cdot\bm{\sigma}\hat{G}({\color{black}i\omega_\nu},\bm{k}')\bm{M}_{\bm{k}'-\bm{k}''}\cdot\bm{\sigma}\hat{G}({\color{black}i\omega_\nu},\bm{k}'')\bm{M}_{\bm{k}''-\bm{k}'''}\bm{\sigma}\hat{G}\left({\color{black}i\omega_\nu},\bm{k}'''\right)\bm{M}_{\bm{k}'''-\bm{k}}\cdot\bm{\sigma}\Big\}\no\\
&=&\frac{\tilde{\beta}}{2}\left(\left|\bm{M}_1\right|^2+\left|\bm{M}_2\right|^2\right)^2+\frac{\beta-\tilde{\beta}}{2}\left(\left|\bm{M}^2_1\right|^2+\left|\bm{M}^2_2\right|^2\right)+\left(g-\tilde{\beta}\right) \left|\bm{M}_1\right|^2\left|\bm{M}_2\right|^2+\frac{\tilde{g}}{2}\left(\left|\bm{M}_{1}\cdot\bm{M}_{2}\right|^2+\left|\bm{M}_{1}\cdot\bm{M}^{*}_{2}\right|^2\right).\quad\qquad
\eea

\noi Once again we start with the calculation of the trace of the projectors: 
\begin{align}
 & A_{\bm{k},s;\bm{k}',s';\bm{k}'',s'';\bm{k}''',s'''}^{\left(4\right)} = \mathrm{Tr}\left\{ \hat{\mathcal{P}}_{s}(\bm{k})\hat{\mathcal{P}}_{s'}\left(\bm{k}'\right)\hat{\mathcal{P}}_{s''}\left(\bm{k}''\right)\hat{\mathcal{P}}_{s'''}\left(\bm{k}'''\right)\right\} \nonumber \\
=& \frac{1}{8}\left(1+ss'\hat{\bm{d}}_{\bm{k}}\cdot\hat{\bm{d}}_{\bm{k}'}+ss''\hat{\bm{d}}_{\bm{k}}\cdot\hat{\bm{d}}_{\bm{k}''}+ss'''\hat{\bm{d}}_{\bm{k}}\cdot\hat{\bm{d}}_{\bm{k}'''}+s's''\hat{\bm{d}}_{\bm{k}'}\cdot\hat{\bm{d}}_{\bm{k}''}\right.+ s's'''\hat{\bm{d}}_{\bm{k}'}\cdot\hat{\bm{d}}_{\bm{k}'''}+s''s'''\hat{\bm{d}}_{\bm{k}''}\cdot\hat{\bm{d}}_{\bm{k}'''}\nonumber \\
&+ i\epsilon_{ijk}ss's''\hat{\bm{d}}_{\bm{k}}^{i}\hat{\bm{d}}_{\bm{k}'}^{j}\hat{\bm{d}}_{\bm{k}''}^{k}
+ i\epsilon_{ijk}ss's'''\hat{\bm{d}}_{\bm{k}}^{i}\hat{\bm{d}}_{\bm{k}'}^{j}\hat{\bm{d}}_{\bm{k}'''}^{k}+i\epsilon_{ijk}ss''s'''\hat{\bm{d}}_{\bm{k}}^{i}\hat{\bm{d}}_{\bm{k}''}^{j}\hat{\bm{d}}_{\bm{k}'''}^{k}+i\epsilon_{ijk}s's''s'''\hat{\bm{d}}_{\bm{k}'}^{i}\hat{\bm{d}}_{\bm{k}''}^{j}\hat{\bm{d}}_{\bm{k}'''}^{k}\nonumber\\
& + \left.\left(\delta_{ij}\delta_{kl}-\delta_{ik}\delta_{jl}+\delta_{il}\delta_{jk}\right)ss's''s'''\hat{\bm{d}}_{\bm{k}}^{i}\hat{\bm{d}}_{\bm{k}'}^{j}\hat{\bm{d}}_{\bm{k}''}^{k}\hat{\bm{d}}_{\bm{k}'''}^{l}\right)\,.
\end{align}

The fourth order free energy contains two type of terms. The one only involves one of the two order parameters $\bm{M}_{1,2}$, while the other involves both. We find the following expressions:
\bea
\frac{\tilde{\beta}}{2}&=&\frac{8T}{N}\sum_{{\color{black}i\omega_\nu},\bm{k}} \sum_{s,s',s'',s'''}^\pm\left(A_{\bm{k},s;\bm{k}+\bm{Q}_{1\left(2\right)},s';\bm{k},s'';\bm{k}+\bm{Q}_{1\left(2\right)},s'''}^{\left(4\right)}G_{{\color{black}i\omega_\nu},\bm{k}}^{s}G_{{\color{black}i\omega_\nu},\bm{k}+\bm{Q}_{1\left(2\right)}}^{s'}G_{{\color{black}i\omega_\nu},\bm{k}}^{s''}G_{{\color{black}i\omega_\nu},\bm{k}+\bm{Q}_{1\left(2\right)}}^{s'''}\right)\,,\no\\
\frac{\beta-\tilde{\beta}}{2}&=&\frac{2T}{N}\sum_{{\color{black}i\omega_\nu},\bm{k}}\sum_{s,s',s'',s'''}^\pm\Bigg\{4A_{\bm{k}+\bm{Q}_{1\left(2\right)},s;\bm{k},s';\bm{k}-\bm{Q}_{1\left(2\right)},s'';\bm{k},s'''}^{(4)}G_{{\color{black}i\omega_\nu},\bm{k}}^{s}G_{{\color{black}i\omega_\nu},\bm{k}+\bm{Q}_{1\left(2\right)}}^{s'}G_{{\color{black}i\omega_\nu},\bm{k}}^{s''}G_{{\color{black}i\omega_\nu},\bm{k}-\bm{Q}_{1\left(2\right)}}^{s'''}\no\\
&&\qquad\quad\qquad\quad-2A_{\bm{k},s;\bm{k}+\bm{Q}_{1\left(2\right)},s';\bm{k},s'';\bm{k}+\bm{Q}_{1\left(2\right)},s'''}^{\left(4\right)}G_{{\color{black}i\omega_\nu},\bm{k}}^{s}G_{{\color{black}i\omega_\nu},\bm{k}+\bm{Q}_{1\left(2\right)}}^{s'}G_{{\color{black}i\omega_\nu},\bm{k}}^{s''}G_{{\color{black}i\omega_\nu},\bm{k}+\bm{Q}_{1\left(2\right)}}^{s'''}\Bigg\}\no\,,\\ 
g&=&\frac{2T}{N}\sum_{{\color{black}i\omega_\nu},\bm{k}}\sum_{s,s',s'',s'''}^\pm\Bigg\{4A_{\bm{k},s;\bm{k}+\bm{Q}_{1},s';\bm{k},s'';\bm{k}+\bm{Q}_{2},s'''}^{\left(4\right)}G_{{\color{black}i\omega_\nu},\bm{k}}^{s}G_{{\color{black}i\omega_\nu},\bm{k}+\bm{Q}_{1}}^{s'}G_{{\color{black}i\omega_\nu},\bm{k}}^{s''}G_{{\color{black}i\omega_\nu},\bm{k}+\bm{Q}_{2}}^{s'''}\no\\
&&\qquad\qquad-A_{\bm{k},s;\bm{k}+\bm{Q}_{1},s';\bm{k}+\bm{Q}_{1}+\bm{Q}_{2},s'';\bm{k}+\bm{Q}_{2},d}^{\left(4\right)}G_{{\color{black}i\omega_\nu},\bm{k}}^{a}G_{{\color{black}i\omega_\nu},\bm{k}+\bm{Q}_{1}}^{s'}G_{{\color{black}i\omega_\nu},\bm{k}+\bm{Q}_{1}+\bm{Q}_{2}}^{s''}G_{{\color{black}i\omega_\nu},\bm{k}+\bm{Q}_{2}}^{s'''}\no\\
&&\qquad\qquad-A_{\bm{k},s;\bm{k}+\bm{Q}_{2},s';\bm{k}+\bm{Q}_{1}+\bm{Q}_{2},s'';\bm{k}+\bm{Q}_{1},d}^{\left(4\right)}G_{{\color{black}i\omega_\nu},\bm{k}}^{a}G_{{\color{black}i\omega_\nu},\bm{k}+\bm{Q}_{2}}^{s'}G_{{\color{black}i\omega_\nu},\bm{k}+\bm{Q}_{1}+\bm{Q}_{2}}^{s''}G_{{\color{black}i\omega_\nu},\bm{k}+\bm{Q}_{1}}^{s'''}\Bigg\},\quad\qquad\no\\
\frac{\tilde{g}}{2}&=&\frac{2T}{N}\sum_{{\color{black}i\omega_\nu},\bm{k}}\sum_{s,s',s'',s'''}^\pm\Bigg\{A_{\bm{k},s;\bm{k}+\bm{Q}_{1},s';\bm{k}+\bm{Q}_{1}+\bm{Q}_{2},s'';\bm{k}+\bm{Q}_{2},s'''}^{\left(4\right)}G_{{\color{black}i\omega_\nu},\bm{k}}^{s}G_{{\color{black}i\omega_\nu},\bm{k}+\bm{Q}_{1}}^{s'}G_{{\color{black}i\omega_\nu},\bm{k}+\bm{Q}_{1}+\bm{Q}_{2}}^{s''}G_{{\color{black}i\omega_\nu},\bm{k}+\bm{Q}_{2}}^{s'''}\no\\
&&\quad\qquad+A_{\bm{k},s;\bm{k}+\bm{Q}_{2},s';\bm{k}+\bm{Q}_{1}+\bm{Q}_{2},s'';\bm{k}+\bm{Q}_{1},s'''}^{\left(4\right)}G_{{\color{black}i\omega_\nu},\bm{k}}^{s}G_{{\color{black}i\omega_\nu},\bm{k}+\bm{Q}_{2}}^{s'}G_{{\color{black}i\omega_\nu},\bm{k}+\bm{Q}_{1}+\bm{Q}_{2}}^{s''}G_{{\color{black}i\omega_\nu},\bm{k}+\bm{Q}_{1}}^{s'''}\Bigg\}\no.
\eea

\noi By using the following relation:
\bea
&&\sum_{s,s^{\prime},s^{\prime\prime},s^{\prime\prime\prime}=\pm} A_{\bm{k},s;\bm{k}+\bm{Q}_{1},s^{\prime};\bm{k}+\bm{Q}_{1}+\bm{Q}_{2},s^{\prime\prime};\bm{k}+\bm{Q}_{2},s^{\prime\prime\prime}}^{\left(4\right)}+A_{\bm{k},s;\bm{k}+\bm{Q}_{2},s^{\prime};\bm{k}+\bm{Q}_{1}+\bm{Q}_{2},s^{\prime\prime};\bm{k}+\bm{Q}_{1},s^{\prime\prime\prime}}^{\left(4\right)}\no\\
&&\qquad\qquad\qquad\qquad=2\Re\left(A_{\bm{k},s;\bm{k}+\bm{Q}_{1},s^{\prime};\bm{k}+\bm{Q}_{1}+\bm{Q}_{2},s^{\prime\prime};\bm{k}+\bm{Q_{2},s^{\prime\prime\prime}}}^{\left(4\right)}\right) 
\eea

\noi the above expressions are simplified as follows:
\bea
g&=&\frac{2T}{N}\sum_{{\color{black}i\omega_\nu},\bm{k}}\sum_{s,s',s'',s'''}^\pm\Bigg\{4A_{\bm{k},s;\bm{k}+\bm{Q}_{1},s';\bm{k},s'';\bm{k}+\bm{Q}_{2},s'''}^{(4)}G_{{\color{black}i\omega_\nu},\bm{k}}^sG_{{\color{black}i\omega_\nu},\bm{k}+\bm{Q}_1}^{s'}G_{{\color{black}i\omega_\nu},\bm{k}}^{s''}G_{{\color{black}i\omega_\nu},\bm{k}+\bm{Q}_2}^{s'''}\no\\
&&\quad-2\Re\left(A_{\bm{k},s;\bm{k}+\bm{Q}_1,s';\bm{k}+\bm{Q}_1+\bm{Q}_2,s'';\bm{k}+\bm{Q}_2,d}^{(4)}\right)G_{{\color{black}i\omega_\nu},\bm{k}}^aG_{{\color{black}i\omega_\nu},\bm{k}+\bm{Q}_1}^{s'}G_{{\color{black}i\omega_\nu},\bm{k}+\bm{Q}_1+\bm{Q}_2}^{s''}G_{{\color{black}i\omega_\nu},\bm{k}+\bm{Q}_2}^{s'''}\Bigg\},\no\\
\frac{\tilde{g}}{2}&=&\frac{2T}{N}\sum_{{\color{black}i\omega_\nu},\bm{k}}\underset{s,s',s'',s'''=\pm}{\sum}2\Re\left(A_{\bm{k},s;\bm{k}+\bm{Q}_1,s';\bm{k}+\bm{Q}_1+\bm{Q}_2,s'';\bm{k}+\bm{Q}_2,s'''}^{\left(4\right)}\right)G_{{\color{black}i\omega_\nu},\bm{k}}^sG_{{\color{black}i\omega_\nu},\bm{k}+\bm{Q}_1}^{s'}G_{{\color{black}i\omega_\nu},\bm{k}+\bm{Q}_1+\bm{Q}_2}^{s''}G_{{\color{black}i\omega_\nu},\bm{k}+\bm{Q}_2}^{s'''}.\no
\eea

\noi The numerical calculation of the above coefficient is performed with a cut-off of $\left|k_x\right|,\left|k_x\right|\leq1.5$ and a mesh of $N=100^2$ points in $\bm{k}$ space. Figure~\ref{fig:FigureApp1} collects our results for the magnetic phase diagrams discussed in the main text.

\begin{figure*}[h!]
\centering
\includegraphics[width=\textwidth]{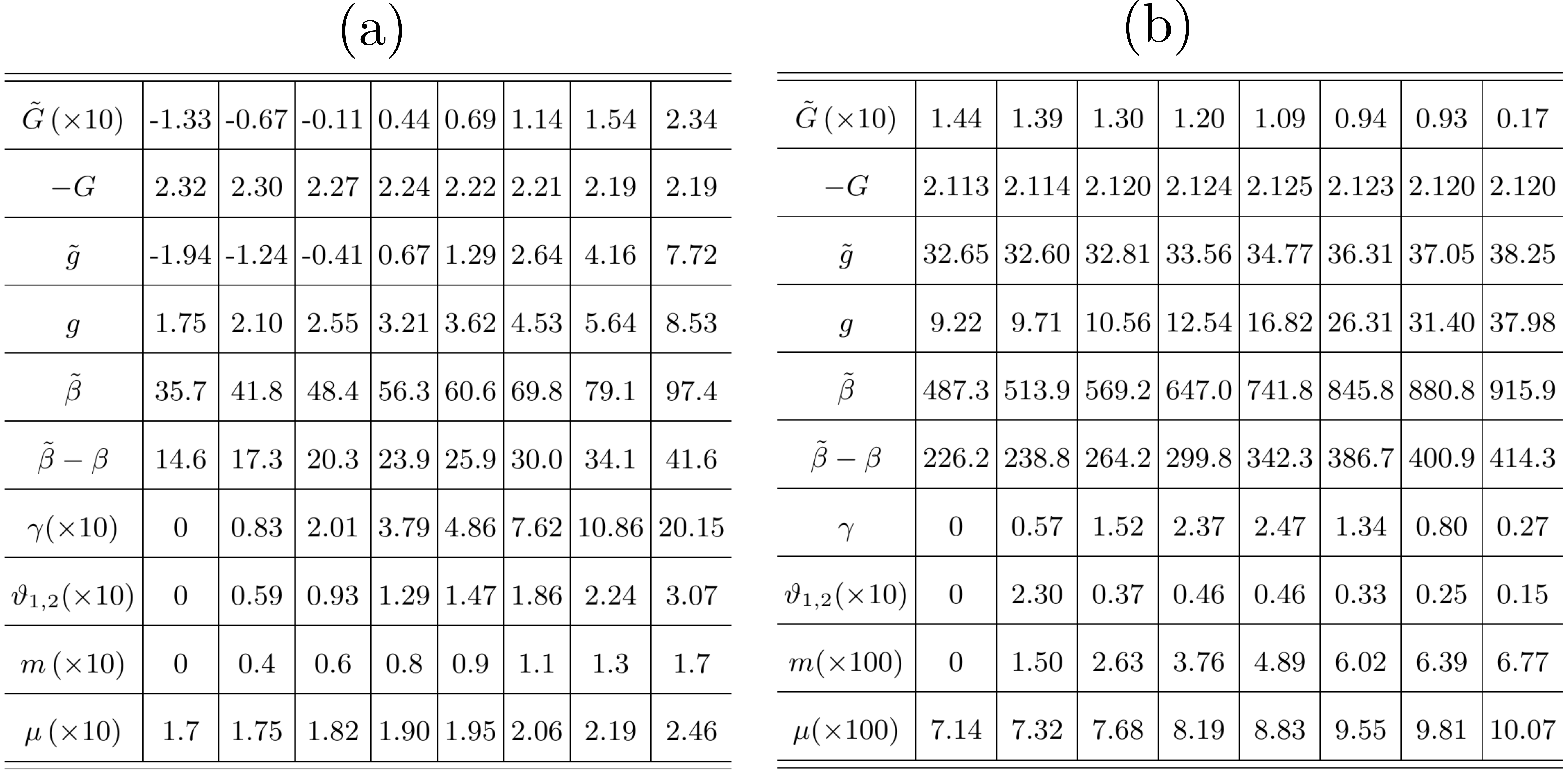}
\caption{(a) The table shows the data employed to obtain Fig.~\ref{fig:Figure2}(a) using the expressions defined in App.~\ref{app:AppendixD}. Each column of the table corresponds, one-to-one from left to right, to a pair of points in Fig.~\ref{fig:Figure2} which are obtained for by discarding and including, respectively, the contribution of $\gamma$ which is induced by a nonzero $m$. Our calculations are carried out for the temperature value $T=1/30$ and the chemical potential is $\mu=0.17$ when $m=0$. The Schr\"odinger masses are set to be $m_1=1$ and $m_2=0.3$. For all our calculations, the chemical potential is suitably adjusted throughout to fix the electron density of the system. Thus the shape of the Fermi surface \PK{stays practically} unchanged and the corresponding nesting vectors are pinned around $\bm{Q}_1\simeq\big(0.7,0\big)$ and $\bm{Q}_2\simeq\big(0,0.7\big)$ with $\chi\big(\bm{Q}_{1,2}\big)\simeq0.98$. To ensure nesting at the above nesting vectors, the che\-mi\-cal potential is suitably adjusted throughout. 
(b) The table shows the data of the corresponding point from right to left in Fig.~\ref{fig:Figure2}(b), upon varying $m$. The calculation is obtained when the temperature is $T=1/70$ and the chemical potential is $\mu=1/14$ when $m=0$. The Schr\"odinger massses are set to be $m_1=1$ and $m_2=0.7$. The chemical potential is tuned to fix the shape of the Fermi surface, and thus the corresponding nesting vectors are pinned around $\bm{Q}_1\sim\big(0.65,0\big)$ and $\bm{Q}_2\sim\big(0,0.65\big)$.}
\label{fig:FigureApp1}
\end{figure*}

\newpage \section{Additional calculations on the robustness of the Sk-SVC chiral topological superconductor}\label{app:AppendixE}

This appendix presents additional numerical calculations in order to investigate the stability of the chiral TSC with ${\cal N}=2$ which emerges for the case of a Sk-SVC phase. Figure~\ref{fig:FigureApp2} (\ref{fig:FigureApp3}) discusses the stability against the addition of an external field (of charge density wave terms).

\begin{figure*}[h!]
\centering
\includegraphics[width=1\textwidth]{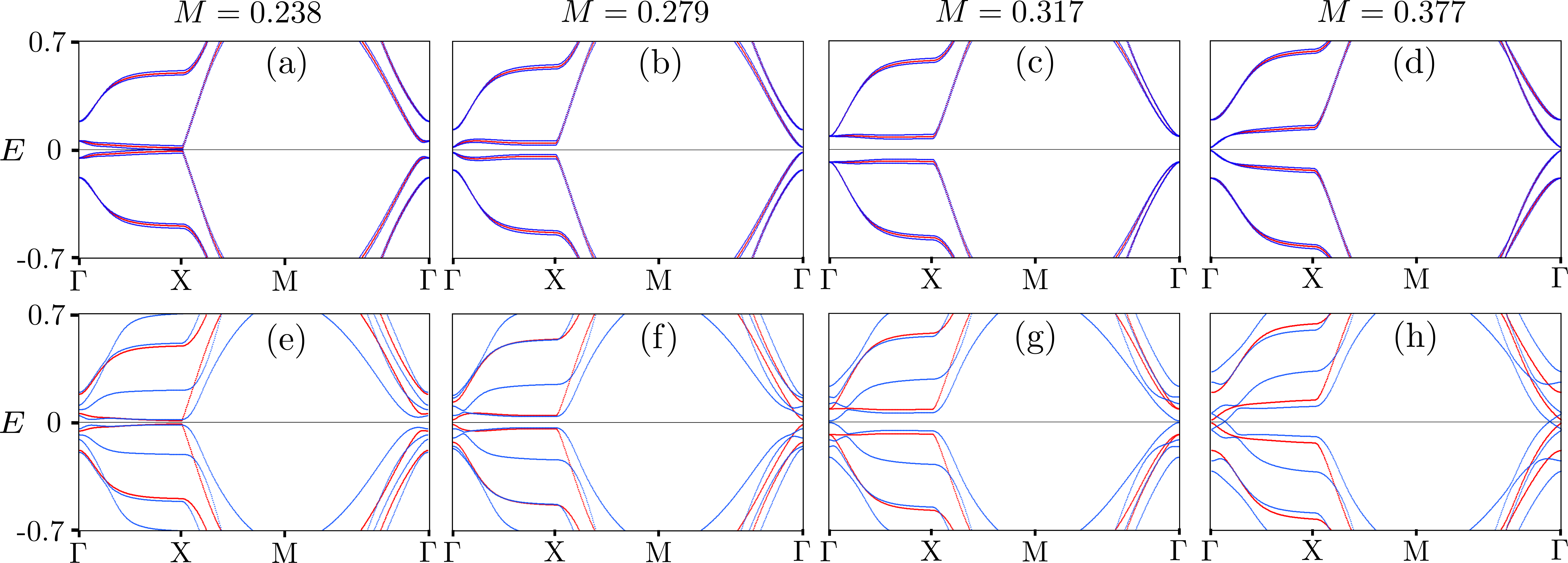}
\caption{Band structure for a Sk-SVC topological superconductor in the additional presence of a Zeeman field $\bm{B}$. The field enters the BdG Hamiltonian through the term $\bm{B}\cdot\bm{\sigma}$. Blue (red) color shows the dispersions in the presence (absence) of the field. (a)-(d) and (e)-(h) show two different cases of field configurations. For the top [bottom] row we have $\bm{B}=0.01(1,\sqrt{2},\sqrt{5})$ [$\bm{B}=0.1(1,\sqrt{2},\sqrt{5})$]. From the above results, we find that the twofold degeneracy becomes generally split but persists at the ${\rm \Gamma}$ point. Hence, the band inversion takes place simultaneously for the two class D Hamiltonian blocks, until the magnetic field strength becomes sufficiently strong to fully split the bands. For the remaining parameters we used the values in Fig.~\ref{fig:Figure5}.}
\label{fig:FigureApp2}
\end{figure*}

\begin{figure*}[h!]
\centering
\includegraphics[width=1\textwidth]{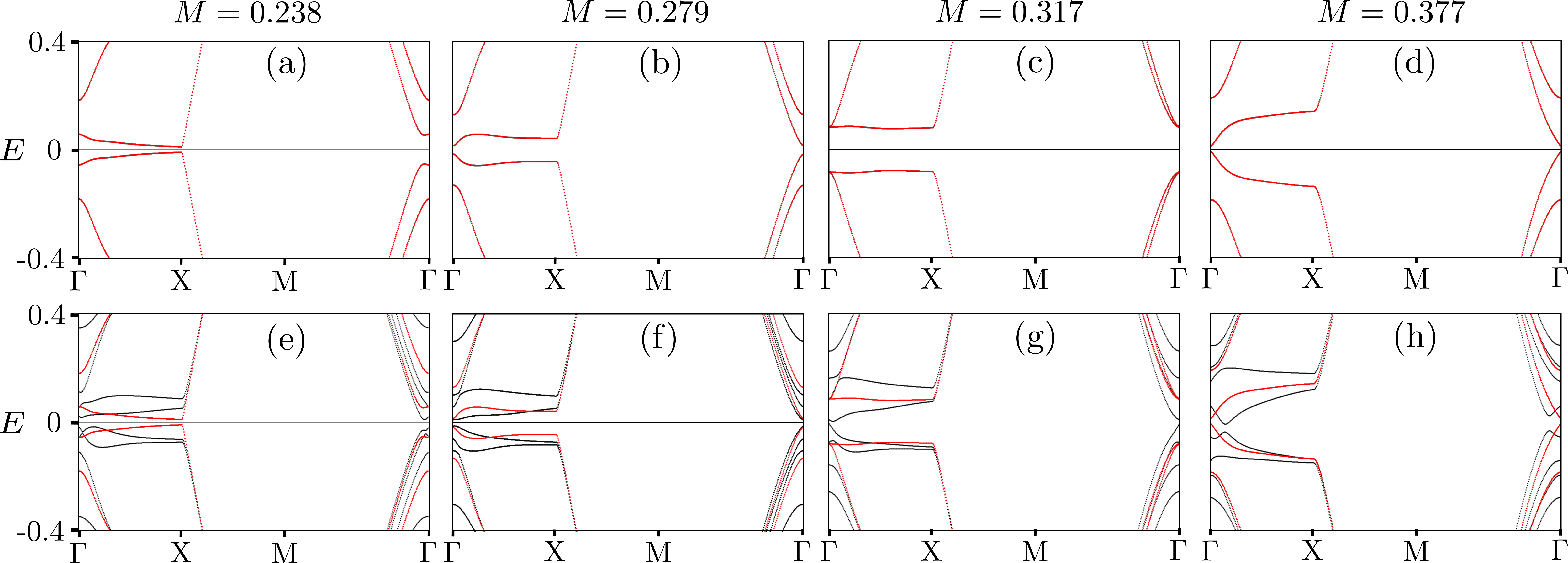}
\caption{Band structure for a Sk-SVC topological superconductor in the additional presence of a combination of charge density wave terms with parameters $W_{\bm{Q}_{1,2,\pm}}\equiv W_{1,2,\pm}$. Each charge density wave component enters the BdG Hamiltonian in a similar fashion to $\bm{M}_{1,2,\pm}\cdot\bm{\sigma}$, albeit that the respective Hamiltonian term is diagonal in spin space and contains a $\tau_3$ matrix instead of $\mathds{1}_\tau$. We consider the configuration where $W_1=We^{-i\pi/5}$, $W_2=We^{-i\pi/4}$, $W_+=We^{-i\pi/3}$ and $ W_-=We^{-i\pi/7}$. Black (red) color shows the dispersions in the presence (absence) of the field. (a)-(d) and (e)-(h) show two different cases for the strength of the charge density waves. For the top (bottom) row $W=0.001$ ($W=0.1$). From the above results, we find that the twofold degeneracy persists at all point for weak values of $W$. However, also here, when $W$ becomes sufficiently strong, it splits the twofold degeneracy in the entire MBZ. For the remaining parameters we used the values considered in Fig.~\ref{fig:Figure5}.}
\label{fig:FigureApp3}
\end{figure*}

\end{widetext}

\end{appendix}

\end{document}